\documentclass{aastex}
\usepackage{emulateapj5}
\usepackage{apjfonts}

\def\wig#1{\mathrel{\hbox{\hbox to 0pt{%
          \lower.5ex\hbox{$\sim$}\hss}\raise.4ex\hbox{$#1$}}}}

\shorttitle{Young Jupiter Atmospheres}
\shortauthors{Fortney, et al.}

\newcommand{\mj}{$M_{\mathrm{J}}$}
\newcommand{\rj}{$R_{\mathrm{J}}$}
\newcommand{\me}{$M_{\oplus}$}

\newcommand{\te}{$T_{\rm eff}$}
\newcommand{\cp}{\citep}
\newcommand{\ct}{\citet}

\begin{document}

\title{Synthetic Spectra and Colors of Young Giant Planet Atmospheres:\\ Effects of Initial Conditions and Atmospheric Metallicity}

\author{J.~J.~Fortney\altaffilmark{1,2}}
\affil{Department of Astronomy and Astrophysics, UCO/Lick Observatory, University of California, Santa Cruz, CA 95064}
\email{jfortney@ucolick.org}
\author{M. S. Marley}
\affil{Space Science and Astrobiology Division, NASA Ames Research Center, Mail Stop 245-3, Moffett Field, CA 94035}
\author{D.~Saumon}
\affil{Los Alamos National Laboratory, PO Box 1663 Mail Stop F663, Los Alamos, NM 87545}
\author{K. Lodders}
\affil{Planetary Chemistry Laboratory, Department of Earth and Planetary Sciences, Washington University, St. Louis, MO 63130}

\altaffiltext{1}{Spitzer Fellow, Space Science and Astrobiology Division, NASA Ames Research Center}
\altaffiltext{2}{Carl Sagan Center, SETI Institute}

\begin{abstract} 

We examine the spectra and infrared colors of the cool methane-dominated atmospheres at \te\ $\le 1400$ K expected for young gas giant planets.  We couple these spectral calculations to an updated version of the Marley et al.~(2007) giant planet thermal evolution models that include formation by core accretion-gas capture.  These relatively cool ``young Jupiters'' can be 1-6 magnitudes fainter than predicted by standard cooling tracks that include a traditional initial condition, which may provide a diagnostic of formation.  If correct, this would make true Jupiter-like planets much more difficult to detect at young ages than previously thought.  Since Jupiter and Saturn are of distinctly super-solar composition, we examine emitted spectra for model planets at both solar metallicity and a metallicity of 5 times solar.  These metal-enhanced young Jupiters have lower pressure photospheres than field brown dwarfs of the same effective temperatures arising from both lower surface gravities and enhanced atmospheric opacity.  We highlight several diagnostics for enhanced metallicity.  A stronger CO absorption band at 4.5 $\mu$m for the warmest objects is predicted.  At all temperatures, enhanced flux in $K$ band is expected due to reduced collisional induced absorption by H$_2$.  This leads to correspondingly redder near infrared colors, which are redder than solar metallicity models with the same surface gravity by up to 0.7 in $J-K$ and 1.5 in $H-K$.  Molecular absorption band depths increase as well, most significantly for the coolest objects.  We also qualitatively assess the changes to emitted spectra due to nonequilibrium chemistry.
\end{abstract}

\keywords{planets and satellites: formation; planetary systems; radiative transfer}

\section{Introduction}
Astronomers around the world are making significant efforts to image planets in orbit around other stars \cp{Beuzit07,Nielsen07,Lafreniere07,Apai08}.  Work on suppressing the glare of potential parent stars has proceeded to the point where contrast ratios of $10^{-5}$ can now typically be achieved on the telescope, and $10^{-7}$ is on the horizon \cp{Macintosh06,Dohlen06}.  Since the contrast ratio for the Jupiter/Sun is $10^{-9}$ \cp[a contrast ratio which has now been achieved in a laboratory,][]{Trauger07}, the majority of this detection work focuses on young stars, as giant planets should be warmest, largest, and brightest when they are young, but will cool, contract, and fade inexorably as they age \cp{Graboske75,Bodenheimer76,Saumon96,Burrows97}.  Given the difficulty of these low contrast ratio observations, the interpretation of observed photometry and spectra takes on great importance.  In practice, when faint planetary candidates are detected, evolution models, which aim to predict the structural and atmospheric properties with age, are needed to convert observed photometry or spectra into a probable planetary mass.

The formation mechanisms of brown dwarfs and giant planets are still not well understood in detail.  While brown dwarfs likely form directly from molecular cloud gas in something akin to the star formation mechanism \cp{Luhman07,Whitworth07}, ``true planets'' form in a disk (IAU definitions aside---for a discussion see Chabrier et al.~2007), perhaps predominantly via core accretion \cp{Lissauer07}.  Recently, discussion has turned to how these distinct formation mechanisms, which may overlap at several Jupiter masses, may leave observational signatures in terms of an object's orbit, evolution, and atmosphere.  A given parent star may well harbor both classes of low mass objects.  This paper addresses the atmospheres of extrasolar giant planets (EGPs), while also describing the spectral properties of our recent work to couple core accretion formation to subsequent planetary evolution \cp{Marley07}.  In this introduction we will first review giant planet evolution models, then discuss our current understanding of the metal-enhanced atmospheres of Jupiter and Saturn.  In \S2 we describe our model atmosphere code, while in \S3 we discuss the differences in atmospheric pressure-temperature (\emph{P-T}) profiles, chemistry, and spectra between models at solar metallicity and those at 5$\times$ solar.  \S4 focuses in particular on the near and mid infrared colors of metal-enhanced atmospheres while in \S5 we discuss and tabulate the near- and mid-IR colors for our ``hot start'' and core-accretion start evolution models.  \S6 addresses nonequilibrium chemistry while \S7 contains additional discussion, caveats, and our conclusions.

\subsection{The Early Evolution of Giant Planets}
Over the past decade only a small number of workers have attempted the difficult task of coupling non-gray radiative-convective atmosphere models to thermal evolution models to enable an understanding of interior structure, atmospheric structure, atmospheric chemistry, and emitted spectra for giant planets and brown dwarfs \cp[e.g.,][]{Burrows97,Chabrier00b,Baraffe03,Saumon08}.  It has perhaps only recently become appreciated by the wider community that these models do not include a mechanism for the formation of the objects that they aim to understand.  The starting point for these models is an arbitrarily large and hot, non-rotating, adiabatic sphere.  These model objects are then allowed to cool and contract from this arbitrary state.  The initial model is soon unimportant, as the cooling and contraction are initially very fast, since the Kelvin-Helmholtz time, $t_{\rm KH}$ is inversely proportional to both luminosity and radius.  Although it is true that the models forget their initial conditions eventually, it is not immediately obvious how long this may take.  In the past, a common thought was that after ``a few million years'' the initial conditions are forgotten and that these standard ``hot start'' evolution models are reliable.  Although this type of model has been successfully applied to Jupiter for decades \cp[e.g.,][]{Graboske75,Hubbard77,Guillot95,FH03} their application to planets at very young ages could potentially be suspect \cp{Stevenson82b}.  More recently \ct{Baraffe02} have investigated similar issues for brown dwarfs.

In order to better understand the properties of gas giant planets at young ages, in \ct{Marley07} we undertook an investigation of the early evolution of giant planets, with initial properties given by a state-of-the-art model of planet formation by core accretion \cp{Hubickyj05}, rather than the traditional (but arbitrary) initial condition which we termed a ``hot start.''  As shown in \mbox{Figure~\ref{quad}}, the post-formation properties of these planets are surprising.  The model planets started their lives smaller and colder than their hot start brethren.  The core accretion start models were less luminous by factors of a few to 100, and the initial conditions were not forgotten for timescales of tens of millions to one billion years.  The reason for the significant difference lies in the treatment of gas accretion \cp[see][]{Marley07}.  In the \ct{Hubickyj05} models the accreting gas arrives at nearly free fall velocity to a shock interface at the protoplanet.  The shock radiation transfer is not followed directly, but a shock jump condition from \ct{Stahler80} is employed; this accretion luminosity is entirely radiated away, leading to the prominent luminosity spike in these models during gas accretion.  The gas that finally accretes onto the planet is therefore relatively cold, low entropy gas.

It is therefore enticing to imagine that one could use the early luminosity, \te, and surface gravity to determine the formation mechanism of a faint planetary-mass companion.  This may be possible, but we caution that the current generation of core accretion formation models \cp{Hubickyj05,Alibert05,Ikoma00} are still only 1D representations of a 3D process.  A detailed look at radiation transfer in the formation shock, as well as incorporating multi-dimensional accretion, should be undertaken before accurate luminosities of young planets can be confidently predicted.  There is, however, another promising avenue for determining ``planethood.''  While a brown dwarf-like companion and its parent star would be expected to share common elemental abundances, the same may not be true of a companion that formed in a disk via core accretion.

\subsection{The Atmospheres of ``True'' Giant Planets}
The \emph{Galileo} and \emph{Cassini} spacecrafts have unambiguously shown us that the atmospheres of Jupiter and Saturn, respectively, are enhanced in heavy elements relative to the Sun.  The \emph{Galileo Entry Probe} measured the abundances of oxygen, carbon, nitrogen, sulfur, and various noble gases in the atmosphere of Jupiter.  Except for oxygen, an enhancement of $\sim$2-4$\times$ solar for each element was found \cp{Atreya03}, although the oxygen abundance determination may have been hindered by meteorological effects \cp{Showman98}.  Saturn's atmosphere is enhanced in carbon by factor of $\sim$10, from an analysis of \emph{Cassini} spectra \cp{Flasar05}, and in phosphorus by a factor of $\sim$7, from \emph{ISO} spectra \cp{deGraauw97,Visscher05}.  If Jupiter, Saturn, and EGPs formed through a common mechanism, we can expect EGP atmospheres to have high metallicities as well.

How this atmospheric metallicity (and indeed the \emph{ratios of specific elements}) may be set, as a function of planet mass, orbital distance, disk mass, disk metallicity, etc., is still open territory.  In particular, the relative importance of processes that have enriched the atmospheres of Jupiter and Saturn is still unclear.  These potentially include planetesimal bombardment and accumulation during formation \cp{Owen99,Gautier01a,Gautier01b,Guillot00,Alibert05b}, erosion of the heavy element core \cp{Stevenson85,Guillot04}, direct accretion of metal-rich disk gas \cp{Guillot06a}, and chemical fractionation within the planet \cp{SS77b,Lodders04}.  Clearly observations of extrasolar giant planet will shed light on giant planet formation.  Here we will undertake a first step at exploring how the spectra and colors of uniformly metal-enhanced atmospheres differ from strictly solar composition.

Since there is some evidence from both observations \cp{Chauvin05} and theory \cp{Boss01b,Kroupa03,Ida04} that the ``planetary'' and ``stellar'' formation modes may overlap at several Jupiter masses, it will be important to be able to decipher a formation mechanism based on observable properties.  Some of this work has progressed on the orbital dynamics of given companions or classes of companions, such as by \ct{Ribas07}, who found different eccentricity distributions for the radial velocity planets above and below 4 \mj.  Additionally, transiting planet mass and radius determinations allow for calculations of bulk planet density, which shed light on the internal abundance of heavy elements \cp{Guillot06,Fortney07a,Burrows07,Baraffe08}.  A hallmark of Jupiter and Saturn is that they are enriched in heavy elements compared to the Sun, which is known from planet structure models \cp{Podolak74,Saumon04}.  These heavy elements are partitioned between a dense core and an enrichment within the H/He envelope \cp{Hubbard89,Guillot99,Saumon04}.  All planets which form in disks around young stars are expected to be enriched in heavy elements due to these disks possessing both abundant gas \emph{and} solids \cp{Pollack96,Ida04}.

Another rewarding pathway for differentiating planets from low mass brown dwarfs is from direct characterization of their atmospheres.  One could spectroscopically measure the heavy element abundances of the H/He envelope directly, as has long been done in the solar system \cp[e.g.][]{Gautier89,Encrenaz05}.  If Nature is able to form two kinds of objects with an overlapping mass distribution, these distinct formation modes may leave distinct observable atmospheric metallicities in the atmospheres of these objects \cp{Chabrier07,Marley07b}.  Therefore, two classes of planetary mass companions may be revealed by their emitted spectra.

There are then two tasks to be completed.  We first will investigate the differences in spectra and infrared colors between models of solar composition (M/H=[0.0]) and those with a metallicity enhanced by a factor of five (M/H=[0.7]).  This metallicity enhancement is similar to that of Jupiter and Saturn, but still far removed from the 30-40$\times$ solar enhancement (at least in carbon) that has been measured for Uranus and Neptune \cp{Gautier89}.  This investigation will be done at the low gravities and effective temperature most relevant for EGPs, (log $g$$\lesssim$4.3, \te$\lesssim$1400 K), whereas field T-type brown dwarfs with similar \te s in general have surface gravities 10 times larger.  Since we restrict ourselves to \te$<$1400 K, we will necessarily be targeting cloud-free CH$_4$-rich, rather than CO-rich, atmospheres.  We focus on these relatively cool objects since a detection of CH$_4$ together with an age estimate would significantly constrain a planetary candidate's mass, even taking into account the uncertainties in evolution models.  Later we will examine the evolution of infrared spectra and colors at 5$\times$ solar metallicity, specifically for the \ct{Marley07} evolution models for giant planets.  As anticipated, the lower \te\ and radii for these models lead to dramatically fainter absolute magnitudes compared to hot start 1$\times$ solar models.

\section{Model Description}
We employ a 1D model atmosphere code that has been used for a variety of planetary and substellar objects.  Recently it has been used for brown dwarfs \citep{Marley02,Saumon06,Saumon07} and EGPs \citep{Fortney05,Fortney06,Marley07b,Fortney07b}.  The radiative transfer method was developed by \citet{Toon89} and has in the past been applied to Titan \cp{Mckay89}, Uranus \cp{MM99}, Gliese 229b \cp{Marley96}, and brown dwarfs in general \cp{Burrows97}.  We use the elemental abundance data of \citet{Lodders03} and compute chemical equilibrium compositions at metallicities of 1$\times$ and 5$\times$ solar, following \citet{Fegley94} and \citet{Lodders02,Lodders06}.  The chemistry calculations include ``rainout,'' where refractory species are depleted from the atmosphere due to their condensation into cloud decks \citep{Lodders99,Burrows99}.  The spectra of brown dwarfs and our solar system's giant planets can only be reproduced when chemistry calculations incorporate this process \cp{Fegley94,Marley02,Burrows02}.  We use the correlated-k method for the tabulation of gaseous opacities \citep{Goody89}; our extensive opacity database is described in \citet*{Freedman08}.  The model atmosphere code is used to compute radiative-convective equilibrium pressure-temperature (\emph{P-T}) profiles and low resolution spectra.  High resolution spectra are computed separately using a full line-by-line radiative transfer code, which utilizes the same chemistry and opacity database.  Since we are modeling warm planets relatively far from their parent stars, here we ignore stellar insolation.

Although we restrict our metal-enhanced planet atmospheres and comparative model spectra to \te$<1400$ K, we have computed 1$\times$ metallicity models up to 2400 K for use in evolutionary calculations.  This is the grid of model atmospheres that serves as the upper boundary condition for modeling the planets' thermal evolution.  Since cloud opacity is predicted to affect the evolution of these planets \cp{Lunine89,Chabrier00b}, we have elected to include it in the evolution atmosphere grid.  We use the cloud model of \ct{AM01} to describe the location, vertical distribution, and particle sizes of major cloud-forming species corundum (Al$_2$O$_3$), iron (Fe), and forsterite (Mg$_2$SiO$_4$).  We assume an $f_{\rm sed}$ sedimentation efficiency parameter of 2, which best matches observations of L dwarfs \cp{Cushing08,Marley08}.  We have generated a grid of cloudy model atmospheres from log $g$=3.0 to 4.5 and \te=500 to 2400 K, supplemented with cloud-free models at \te$<$500 K.  At low \te, the refractory clouds reside very deep in the atmosphere and negligibly effect the spectra and structure \cp{Saumon08}.  A more expansive version of this grid is used in \ct{Saumon08} to compute the thermal evolution of brown dwarfs down to \te$=$500 K.  Our previous planet evolution calculations, presented by \cp{Marley07}, neglected cloud opacity.

The inclusion of cloud opacity into the atmosphere grids leads to some differences in the cooling curves from \mbox{Figure~\ref{quad}}, compared to those from  \ct{Marley07}.  Perhaps most notable is that the core-accretion start planets begin their evolution even colder here, because the cloud opacity closes off an atmospheric radiative zone at pressures of several to tens of bars, which has the effect of leaving a cooler photosphere for a given interior adiabat.  In general, radiative zones can form because the relevant atmospheric opacities, particularly those of water and the pressure-induced opacity of $\rm H_2$, are strongly wavelength-dependent, and opacity windows can appear around 1 to $2\,\rm \mu m$  at temperatures of 1000 to to $2,000\,\rm K$.  The overlap of the Planck function with these windows can allow local radiative transport of energy whereby the temperature profile becomes less steep than an adiabat.  As the local temperature continues to fall and the Planck function moves red-ward still, opacities increase again, closing the radiative window and the local temperature profile can again return to an adiabat.  Such a two-layered convective structure was predicted for Jupiter at pressures of several kilobar and temperatures near 2000 K by \ct{Guillot94a}.  The same effect was subsequently seen in models of brown dwarfs and warm giant planet atmospheres \cp{Marley96,Burrows97,Allard01}, but since such objects are warmer than Jupiter the corresponding radiative region is higher in the atmosphere at lower pressure\footnote{Jupiter's deep radiative window at $1\,\rm \mu m$ originally noted by \ct{Guillot94a} was later found to likely be closed by highly pressure-broadened alkali opacity \cp{Guillot04,Freedman08}.  The potential radiative region in the young giant planet models arises from the longer-wavelength near-infrared windows in water opacity ($JHK$) which are not as strongly affected by alkali opacity. At depth in Jupiter the water opacity windows are closed by the strong pressure-induced opacity of $\rm H_2$ at kilobar pressures.}.

Details of the calculation of the planetary evolution models can be found in \ct{Marley07}.  All planets are assumed to be composed of pure H/He envelopes, with $Y=0.234$ \cp[the value used in][]{Hubickyj05} overlaying a dense olivine core.  The core masses range from 16 \me~(for the 1 \mj~planet) to 19 \me~(for the 10 \mj~planet).

\section{Atmospheres and Spectra at Supersolar Metallicity}
\begin{figure}
\epsscale{0.9}
\plotone{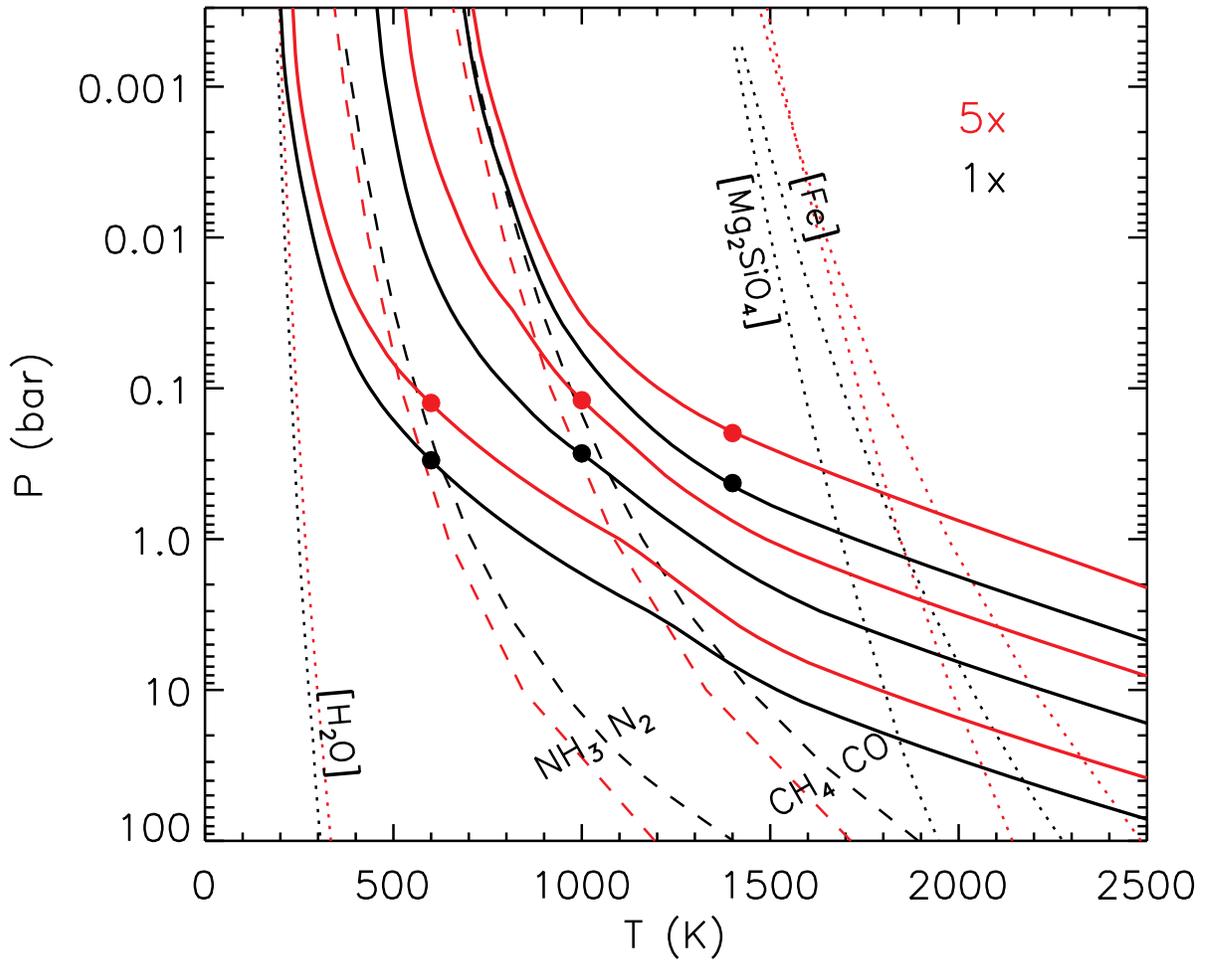}
\caption{Cloud-free \emph{P-T} profiles at 1400, 1000, 600 K at log $g$=3.67.  Curves in black are for 1$\times$ solar metallicity.  Curves in red are for 5$\times$ solar metallicity.  Filled circles indicate the pressure of the mean photosphere, where $T$=\te.  Dotted curves show locations of cloud condensation while dashed curves are chemical equal-abundance boundaries.  Only the 1$\times$ boundaries (black) are labeled.  Note that condensation curves shift to higher temperatures as metallicity increases, while equal-abundance boundaries shift to lower temperatures.
\label{pt1}}
\end{figure}

Before we generate spectra and colors for these evolutionary models, it is first worthwhile to examine the \emph{P-T} conditions for these cool planetary atmospheres.  In \mbox{Figure~\ref{pt1}} are shown a collection of cloud-free \emph{P-T} profiles at a surface gravity of log $g$ (cgs)=3.67, representative of a young 4 \mj\ planet.  Solar composition profiles at 1400, 1000, 600 K are shown in black, while profiles with 5$\times$ solar metallicity are shown in red.  Solid dots indicate where the local temperature equals the \te, illustrating the mean photospheric pressure.  Chemical boundaries are shown in dashed curves and cloud condensation boundaries are dotted curves, using the same black/red color scheme.

Quite clearly, high metallicity atmospheres are everywhere warmer at a given \te.  The higher gaseous opacity of these atmospheres leads to a photospheric pressure that is necessarily lower at a given \te; one cannot see as deeply into a higher opacity atmosphere \cp[e.g.,][]{Saumon94}.  This difference in pressure is a factor of $\sim$2-2.5 for these 5$\times$ solar atmospheres.  This lower photospheric pressure for higher metallicity atmospheres can manifest itself in \emph{lower} contributions to the gaseous opacity for some species, specifically those that are pressure dependent.  These include collisional induced absorption (CIA) by H$_2$ molecules \cp[][where the opacity per unit volume is dependent on the square of the pressure]{Borysow02}, and Na and K, which have strongly pressure broadened optical absorption lines \cp{BMS, Allard03}.  As has been shown in detail elsewhere \cp[e.g.,][]{Lodders02}, condensation and chemical boundaries are non-trivial functions of metallicity.  The higher the metallicity, the higher the temperature at which initial condensation will begin.  However, the CO/CH$_4$ and N$_2$/NH$_3$ equal-abundance curves move to cooler temperatures with increased metallicity \cp{Lodders02}.

The spectra of the profiles plotted in \mbox{Figure~\ref{pt1}} should show strong absorption due to CH$_4$ and H$_2$O, with CO also appearing in the warmer profiles and NH$_3$ in the cooler profiles.  Spectra for a collection of models are shown in \mbox{Figure~\ref{spec1}}.  The spectra for solar metallicity models are shown in black, while those for the 5$\times$ solar models are shown in colors.  We can investigate these spectra for diagnostics of high metallicity.

The most prominent difference in spectra between the ``1$\times$'' and ``5$\times$'' models is a flux enhancement in $K$ band for all 5$\times$ models.  This brightened $K$ band peak was previously shown for a ``5$\times$'' solar model at \te=1200 K by \cp{Chabrier07} in their discussion on differentiating planets from brown dwarfs.  This flux enhancement is tied to the CIA opacity of H$_2$, as it is substantially larger in $K$ band compared to other near infrared wavelengths.  For lower pressures photospheric pressures this CIA absorption is weaker, letting more flux escape in $K$ band, relative to other wavelengths.  Also apparent is the increased CO absorption at 4.5 $\mu$m.  However, the greater CO absorption at 2.3 $\mu$m is swamped by the higher $K$-band flux.  At higher metallicity the abundances of metal-metal species such as CO are increased to a larger degree than those of hydride species.  This leads to stronger CO absorption, relative to absorption bands from other molecules.  This 4.5 $\mu$m CO absorption band will likely be a valuable diagnostic for the warmest young Jupiters \cp{Chabrier07}, but its importance necessarily wanes as the CO abundance drops dramatically below \te$\sim$1000 K for these objects.  Absorption band depths due to H$_2$O and CH$_4$ are modestly deeper at higher metallicity as well.  At the coolest \te s the bands are so deep that it may well be difficult to see any emitted flux, making band depths a difficult diagnostic in practice.

\begin{figure}
%If possible, this figure should large --> wider than just one column
\includegraphics[scale=0.90]{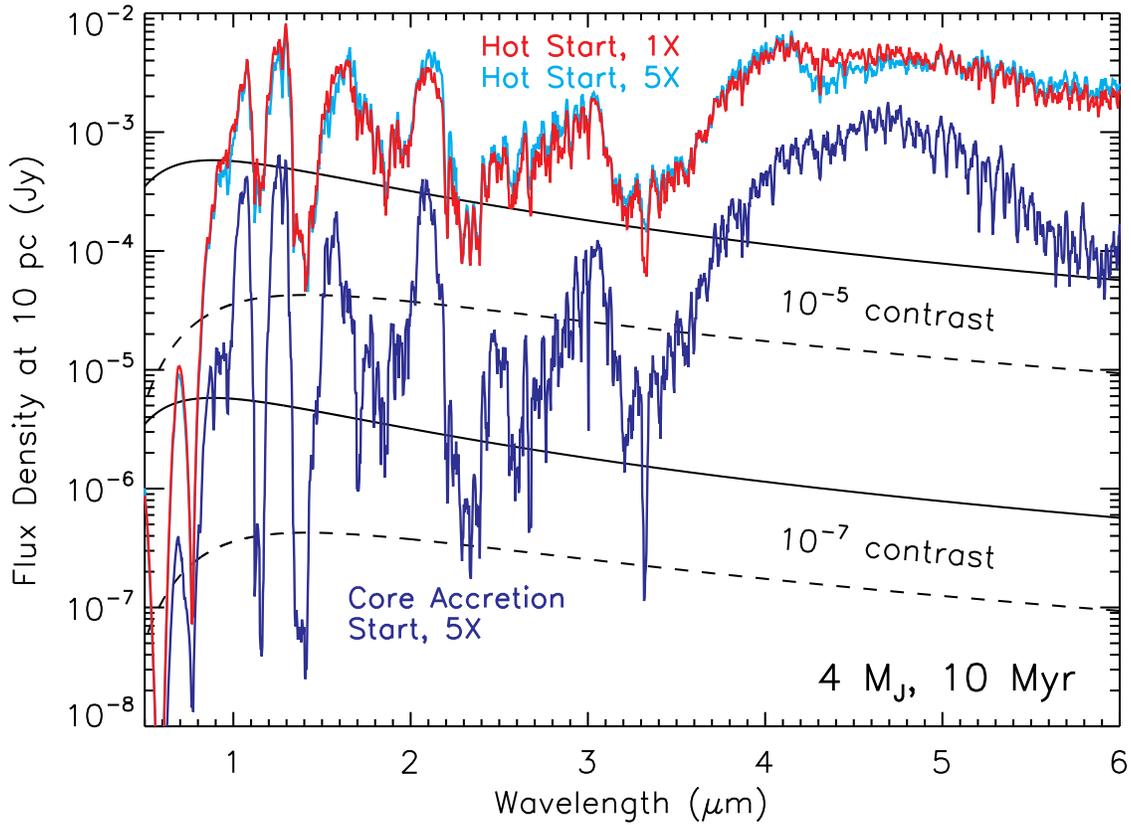}
\caption{Flux density at 10 pc for 4 \mj\ objects at an age of $\sim$10 Myr.  In red is a hot start evolution model with solar metallicity at 1000 K.  In light blue is this same model with 5$\times$ solar metallicity, for comparison.  In dark blue is a 600 K model that uses the core-accretion initial condition and 5$\times$ solar metallicity.  Over-plotted in black are $10^{-5}$ and $10^{-7}$ contrast ratios relative to two blackbody stars.  The two solid curves are for a Sun-like 5770 K star and the dashed curves are for an M2V-like 3600 K star.
\label{spec2}}
\end{figure}

The contrast ratios that will be needed to directly image candidate planets are show in \mbox{Figure~\ref{spec2}}.  The flux density at 10 pc is plotted for three models of a 4 \mj\ planet at an age of $\sim$10 Myr.  Two models utilize the hot start initial condition (and allow a comparison of the effects of metallicity only) while one uses the core accretion initial condition and also has a 5$\times$ solar atmosphere.  Solid curves show the necessary contrast ratios around a Sun-like star, while dashed curves are contrast ratios around an M dwarf.  Strikingly, while the hot start models would be easily detectable at 10$^{-5}$ contrast in the near infrared, the core-accretion start model would be undetectable.  As mentioned above, the H$_2$O and CH$_4$ bands are exceedingly dark, especially for the cooler high metallicity planets, such that contrast ratios of $10^{-9}$ may be necessary to see flux from inside these bands.

\begin{figure}
%If possible, this figure should large --> wider than just one column
\includegraphics[scale=0.90]{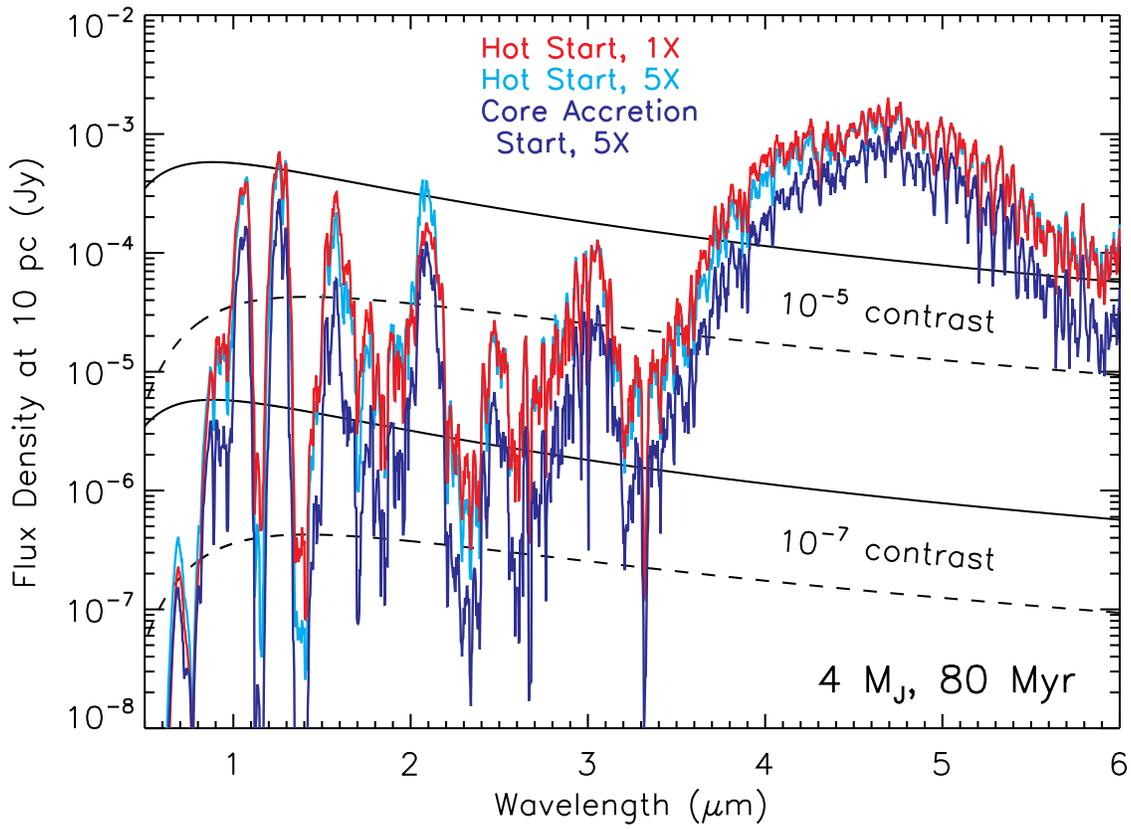}
\caption{This is the same as \mbox{Figure~\ref{spec2}}, on the same scale, but now at age of 80 Myr.  The hot start model has cooled to $\sim$600 K, and the core accretion model to $\sim$500 K.
\label{spec3}}
\end{figure}

\mbox{Figure~\ref{spec3}} allows one to look forward in time to examine these planets at age of 80 Myr.  By this age the two different cooling tracks have nearly run together, meaning the core-accretion model has nearly forgotten its initial condition.  Although the contrast ratios for all bands in the near infrared are smaller for all the models, compared to 10 Myr, one now has the important and distinct advantage that the uncertainties due to the formation history are much smaller.  This will enable a more realistic mass estimate for the planet once one obtains photometric or spectral data.

\begin{figure}
\epsscale{0.9}
\plotone{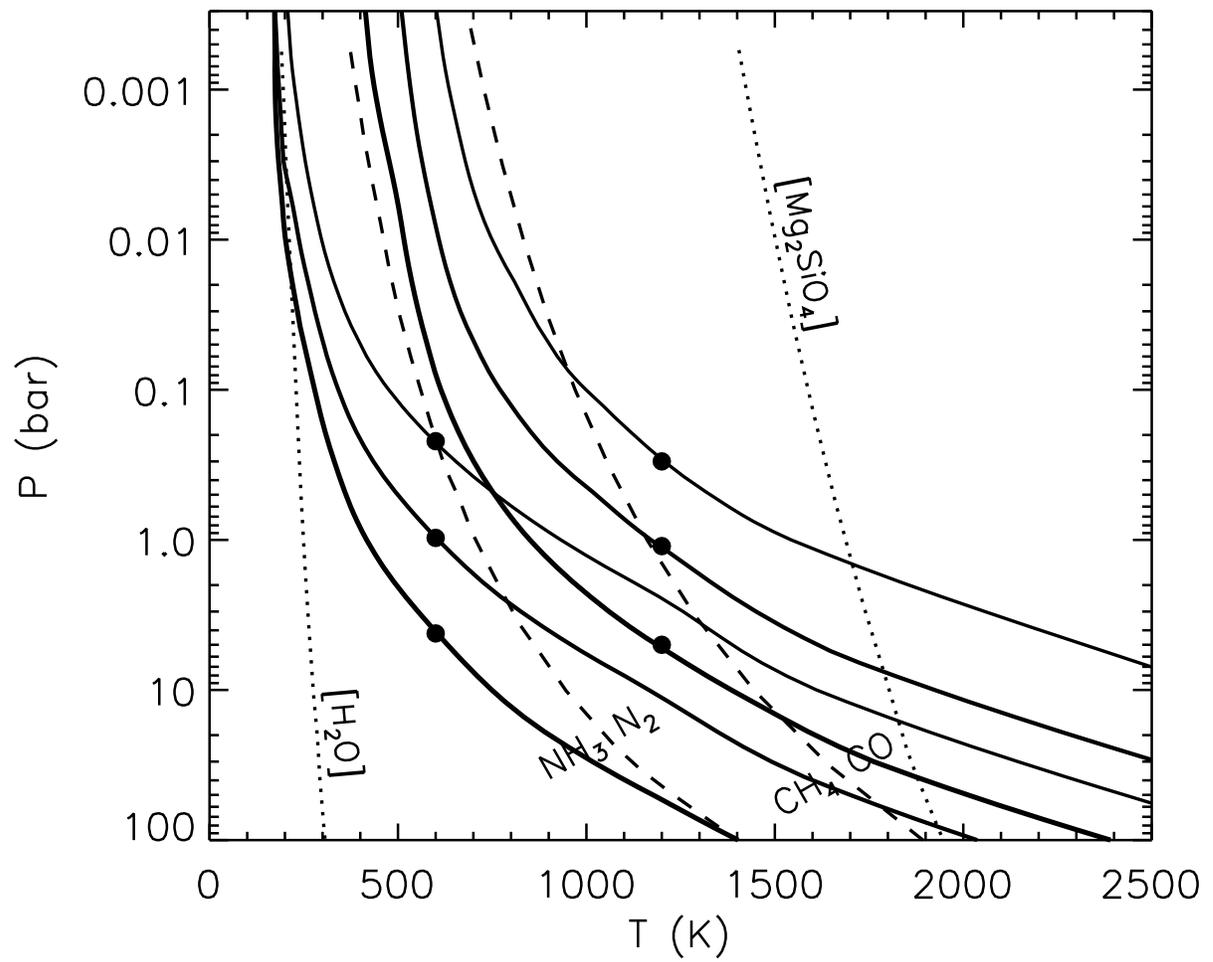}
\caption{Solar metallicity \emph{P-T} profiles at 1200 and 600 K at log $g$=5.5, 4.5, 3.5 (thick to thin lines).
\label{ptg}}
\end{figure}

It is also well known that surface gravity changes photospheric pressures in substellar and planetary objects \cp[e.g.,][]{Burrows97,Burrows02,Kirkpatrick07}.  Since a given optical depth is proportional to column density (as long as CIA opacity does not dominate), this optical depth is reached at a higher pressure in high gravity objects, meaning they will have high pressure photospheres.  This effect is illustrated in \mbox{Figure~\ref{ptg}} for solar metallicity atmospheres at 1200 and 600 K, at gravities of log $g$=3.5, 4.5, and 5.5.  Whereas old massive brown dwarfs commonly have gravities from 5.0 to 5.5, young Jupiters likely exhibit gravities from 3.5 to 4.0, meaning they will have warmer atmospheres at a given \te.  Admittedly, \mbox{Figure~\ref{ptg}} plots relative extremes in gravity, but the large differences in photospheric pressures are significant.  As has been shown by other authors \cp[e.g.,][]{Burrows03b} lower gravity atmospheres must cool to lower \te\ values before crossing a given chemical or condensation boundary.  While the log $g$=5.5 model has clearly passes the water cloud condensation curve at 600 K, the log $g$=3.5 model does not.

It is well understood that lower gravity and higher metallicity both lead to lower pressure photospheres.  Since young Jupiter atmospheres couple both higher metallicity and lower surface gravity it may well be that the visible effects of, for instance, the L-to-T transition, disequilibrium chemistry due to vertical mixing, and water cloud condensation may differ in details compared to high gravity brown dwarfs.  In \S6 we qualitatively address the effects of vertical mixing on carbon chemistry.  Differences in particles sizes, opacity, and vertical extent of clouds are possible, even likely.  Simple cloud models predict cloud particles sizes than may be $\sim$10 times larger in these low gravity environments \citep{AM01,Cooper03} which may lessen the opacity of cloud decks.  While it is still not clear what causes the rapid drop in the opacity of the silicate and iron clouds at the L-to-T transition in brown dwarfs, it should be kept in mind that this transition could be different in character at surface gravities up to 100 times smaller.  For instance, there is already some evidence that the L-to-T transition may occur at lower \te\ in low gravity brown dwarfs \cp{Metchev06,Luhman07b}, which is reasonable since lower gravity atmospheres trace the \emph{P-T} space of higher gravity objects at a lower \te\ \cp{Burrows02,Knapp04}.

\section{``Metallicity Color":  Quantifying Metallicity Effects in Infrared Bands}
\begin{figure}
%If possible, this figure should large --> wider than just one column
\includegraphics[scale=0.78,angle=90]{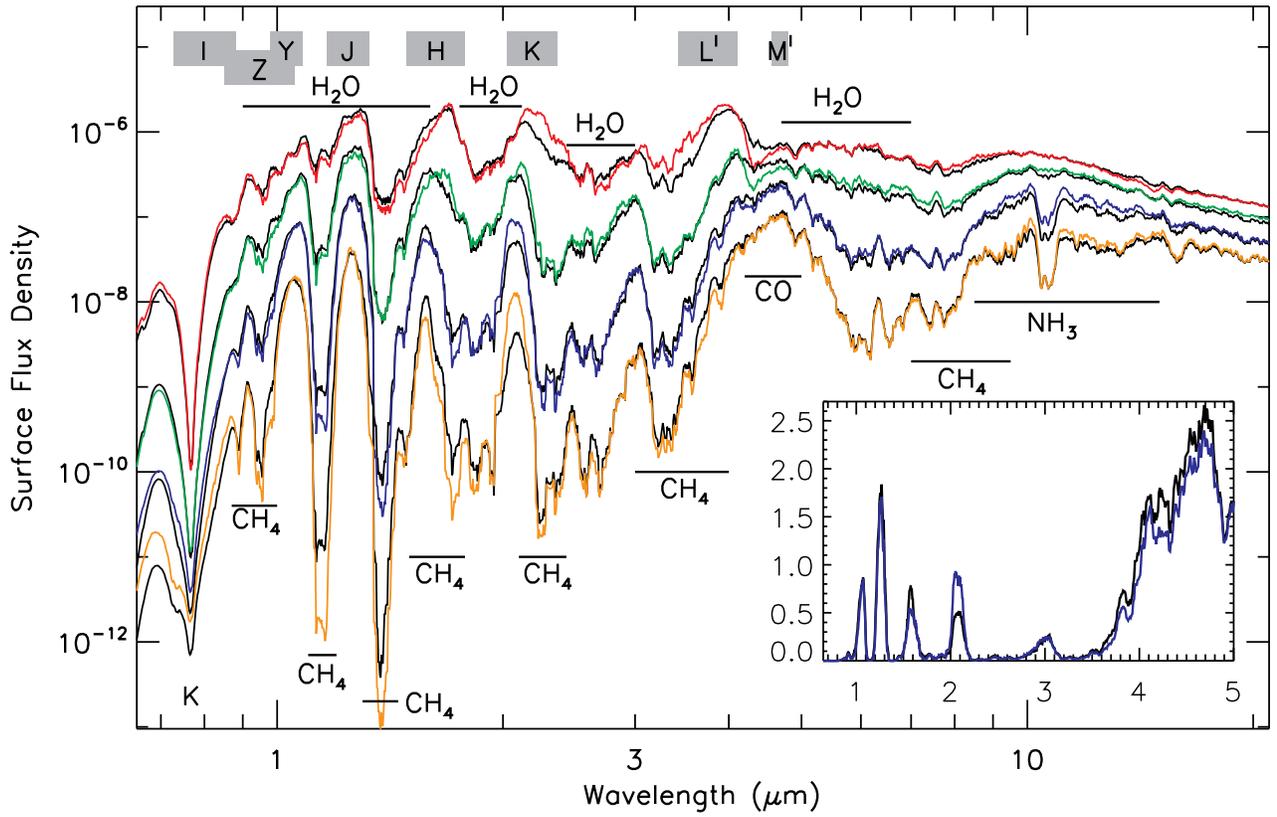}
\caption{Emergent spectra (erg s$^{-1}$ cm$^{-2}$ Hz$^{-1}$) for 5$\times$ solar ([M/H=0.7], colors) and 1$\times$ solar models ([M/H=0.0], black) at, from top to bottom, 1400, 1000, 700, and 500 K, for log $g$=3.67.  The inset shows the 700 K models on a linear $x$ scale (from 0.65 to 5 $\mu$m) and linear $y$ scale ($\times 10^{-7}$, relative to the rest of the figure).
\label{spec1}}
\end{figure}

In \mbox{Figure~\ref{spec1}} we showed spectra of a subset of our 1$\times$ and 5$\times$ solar models.  We can further investigate the differences between these models in the commonly used near and mid IR bands.  Independent of an evolution model (since these models are uncertain at young ages) we can compare the \emph{differences} in magnitudes at various values of \te\ and log $g$, assuming no difference in radii between 1$\times$ and 5$\times$ solar models\footnote{However, a planet with a 5$\times$ solar abundance of metals mixed throughout the entire H/He envelope would be modestly smaller.  The effect of a ``solar'' amount of envelope metals on the structure of brown dwarfs is commonly ignored \cp{Chabrier00,Saumon08}.}.  Here we will term the difference in magnitude between the 5$\times$ and 1$\times$ solar models the ``metallicity color,'' MC=m$_{\rm B5}$-m$_{\rm B1}$ where the letter subscript B is the band and the numerical subscript is the metallicity.  A negative MC means that flux is enhanced in the higher metallicity model.

\mbox{Figure~\ref{mc}} shows the MC as function of \te\ at three values of log $g$ that span the gravity values for the models shown in \mbox{Figure~\ref{mc}}.  Although this is an unusual way to plot colors, it allows for a clear determination of the best diagnostics for high metallicity.  From 1400 K down to 800 K, $J-K$ colors are $\sim$0.7 redder for 5$\times$ models than for 1$\times$ models.  Below 700 K higher metallicity depresses $H$ band flux and even more greatly enhances $K$ band flux, leading to $H-K$ colors becoming even redder than the $J-K$ colors, with $H-K$ high metallicity models becoming redder by 2.0-2.5 than solar composition models, although water cloud condensation below 500 K \citep{Marley02,Burrows03b} could modify this steep falloff.  The $L^{\prime}$ band is interesting in that above 800-900 K flux is enhanced in $L^{\prime}$ at increased metallicity (see \mbox{Figure~\ref{spec1}}) but the flux is depressed in $L^{\prime}$ for the high metallicity models below \te$\sim$800 K.  The remaining bands, $Y$, $Z$, and $M^{\prime}$ are marginally dimmer in the higher metallicity models, with a flux depression that is monotonically weaker for bands $Y$ and $Z$ as gravity increases.

\section{Magnitudes and Colors for Young Jupiter Evolution Models}
\begin{figure}
%If possible, this figure should large --> wider than just one column
\includegraphics[scale=0.78,angle=90]{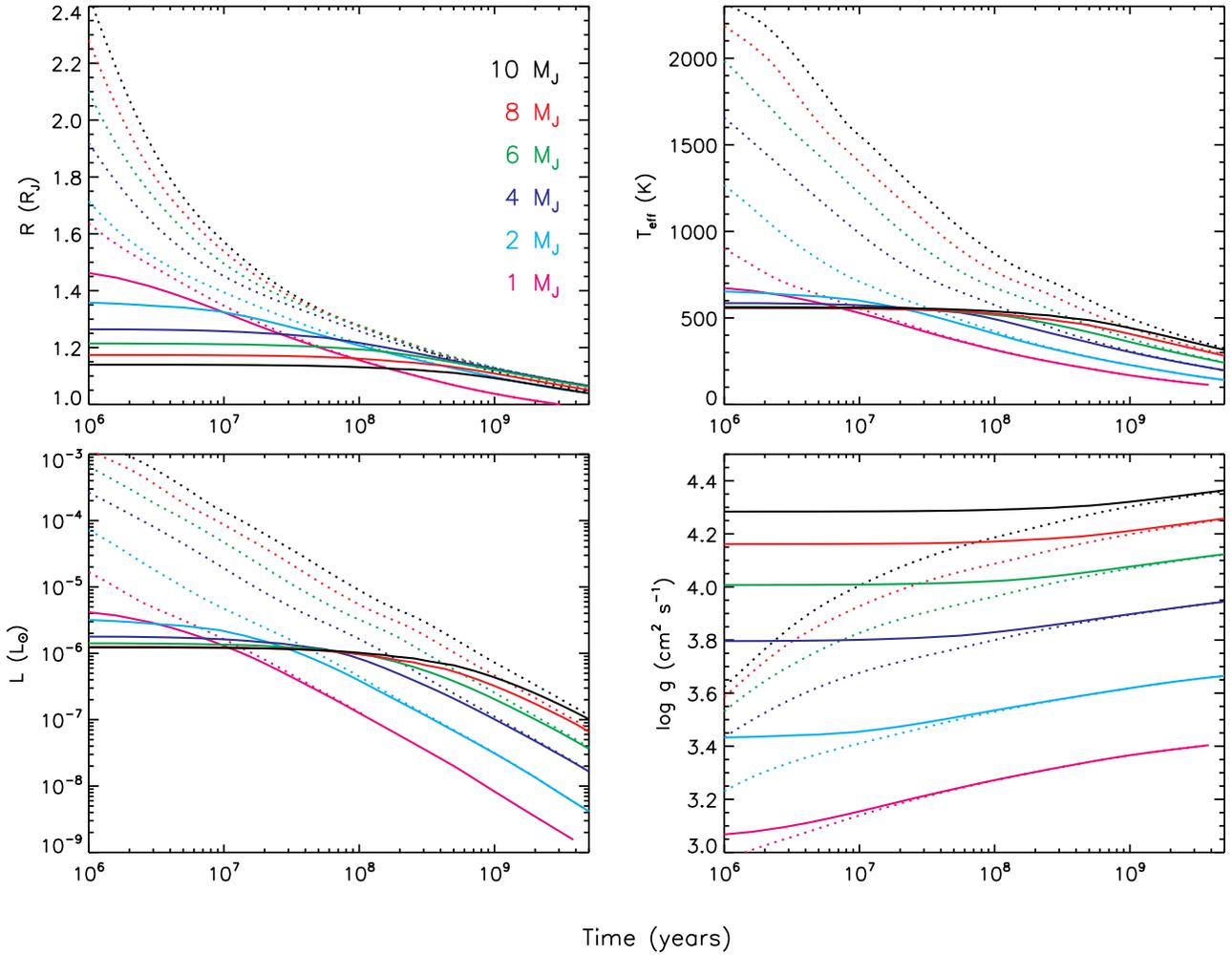}
\caption{Planetary thermal evolution models, updated from \ct{Marley07}.  Dotted lines indicate ``hot start'' planets with an arbitrary initial condition.  Solid lines indicate planets with an initial model from the \ct{Hubickyj05} core accretion formation model.  The model atmosphere grid is 1$\times$ solar and includes the opacity of refractory cloud species.  As in \ct{Marley07}, times on the x-axis are years since formation, which takes no time (by definition) for hot start planets, and $\sim$2.3-3.0 Myr for core accretion planets.
\label{quad}}
\end{figure}

\begin{figure}
\epsscale{0.9}
\plotone{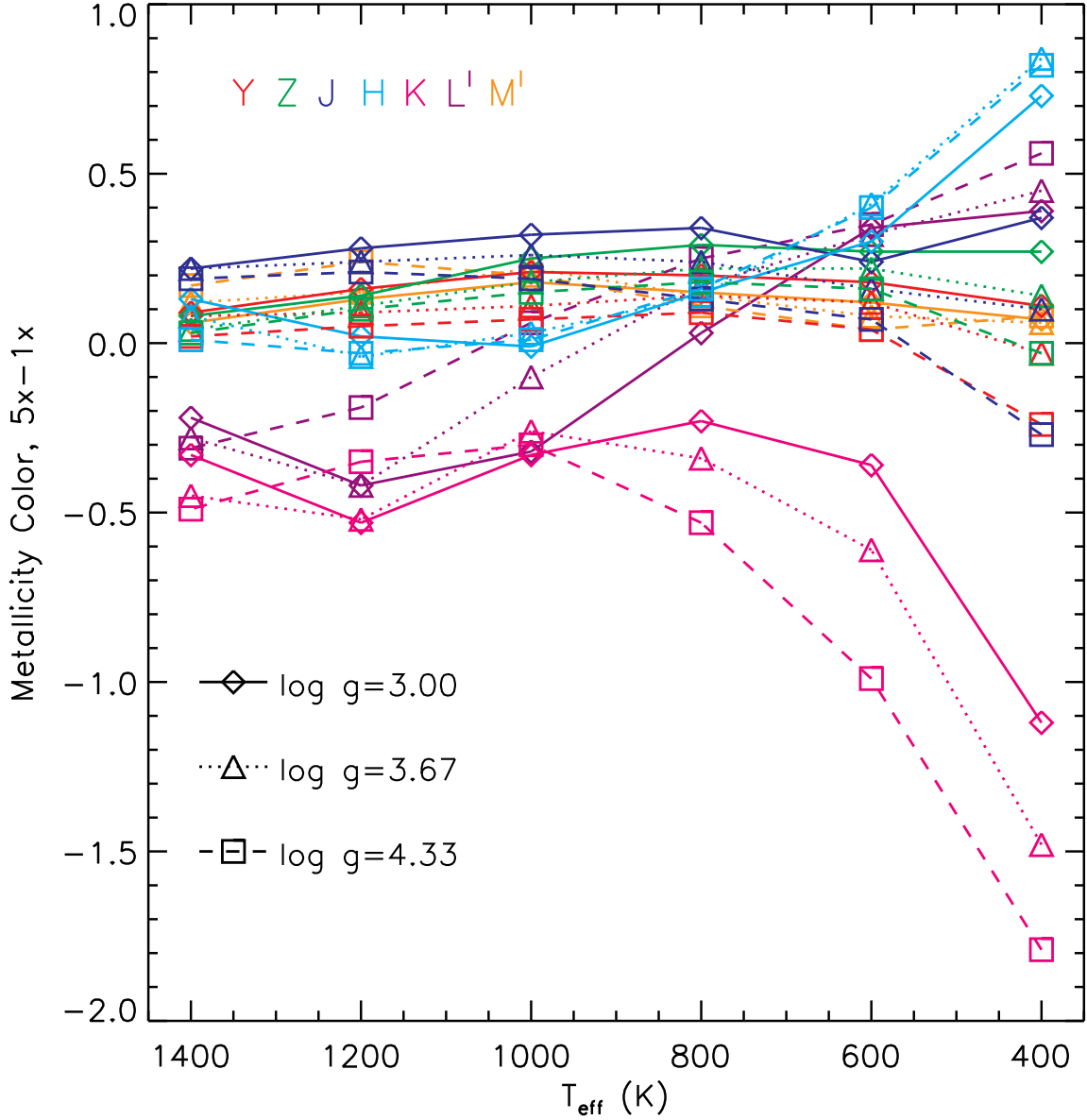}
\caption{Difference in magnitude in a given band (``metallicity color") as a function of \te\ at three surface gravities that span the range of young Jupiter surface gravities shown in \mbox{Figure~\ref{quad}}.  Metallicity color is determined by subtracting the magnitude of the 1$\times$ model from the 5$\times$ model.  Implicit is the assumption that both planets have the same radius.  For instance, at log $g$=3.67 and \te=1000 K, the 5$\times$ solar model is redder in J-K by 0.55 (0.25-(-0.30)). 
\label{mc}}
\end{figure}

We have coupled our spectral models to our updated \ct{Marley07} evolution models show in \mbox{Figure~\ref{quad}}.  This allows for a determination of absolute magnitudes for these model planets as a function of age.  In \mbox{Table~\ref{bigt1}} we provide a table of \te, radius, as well as absolute magnitudes in the standard red-optical and near infrared filters for the hot start 1$\times$ models, which include cloud opacity.  In \mbox{Table~\ref{bigt5}} we provide a similar table for the core-accretion start models, assuming cloud-free 5$\times$ solar metallicity.  This coupling for the core-accretion start models is not strictly self-consistent as the original evolution models used 1$\times$ atmospheres, but this difference is a small one compared to the uncertainty in the initial condition.  Since initial \te s are all below $\sim$800 K, the core-accretion start models span a much smaller range of \te\ than our own hot start models and those of other authors \citep[e.g.,][]{Burrows97,Baraffe03}.  Here we also limit our calculations to \te\ $>$ 400 K, since we also ignore opacity due to water clouds, which condense around \te =400-500 K for these objects \cp{Marley02,Burrows03b}.  We note that \citet{Burrows03b} have shown that water cloud opacity has a relatively modest effect on the spectra of very cool objects.  We will also address this issue in a forthcoming work on the spectra of ultracool dwarfs and planets below 700 K.

\begin{figure}
\epsscale{0.9}
\plotone{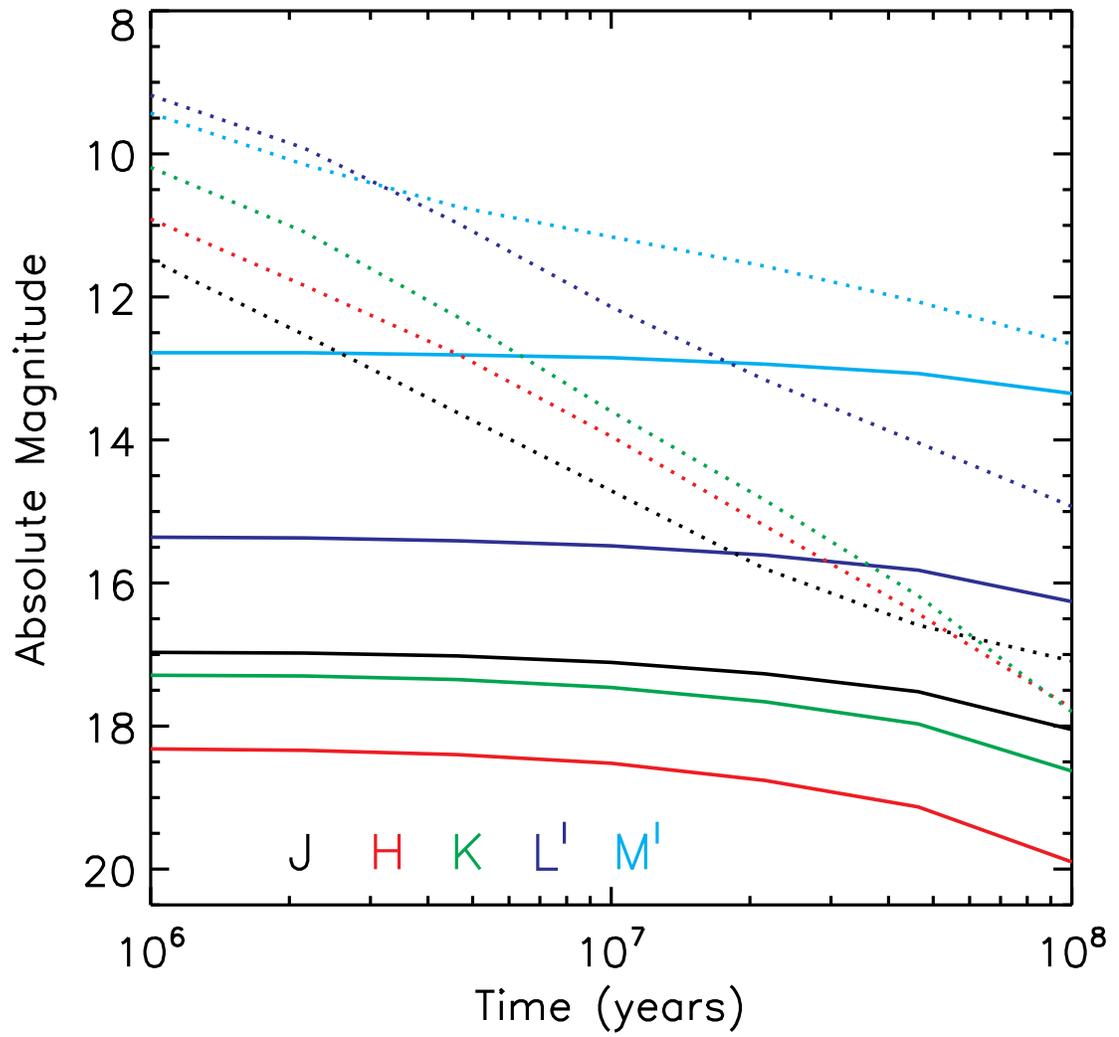}
\caption{Absolute magnitudes vs.~time for a 4 \mj\ planet.  In solid lines is the 5$\times$ core-accretion start model while in dashed lines is the 1$\times$ arbitrary hot start models.
\label{4mj}}
\end{figure}

In \citet{Marley07} and \citet{Fortney05b} we showed that evolution models that incorporate a core accretion initial condition are significantly less luminous that our own ``hot start'' models and those of \citet{Baraffe03} and \citet{Burrows97}.  For example, the luminosity difference for a 4 \mj\ planet is initially a factor of $\sim$100 at 1 Myr.  (See \mbox{Figure~\ref{quad}}.)  Given this factor of 100, one would expect a difference in absolute magnitude of five.  This is indeed born out, as is shown in \mbox{Figure~\ref{4mj}} which compares absolute magnitudes for a 4 \mj\ planet from our hot start 1$\times$ and core-accretion start 5$\times$ calculations.  As indicated in \mbox{Figure~\ref{quad}}, even by $\sim$100 Myr the hot start and core accretion cooling tracks have not yet merged.  This behavior is apparent as well in \mbox{Figure~\ref{4mj}}, which show $\sim$0.5-1.5 magnitude differences at 100 Myr.

\section{Effects of Non-Equilibrium Chemistry}
Another important detection issue is in which bands searches for young Jupiters would be most efficiently carried out.  From 1400 K down to 500 K, the peak in planetary flux gradually shifts from $L^{\prime}$ to $M^{\prime}$ as CO is lost to CH$_4$ \cp{Burrows97}.  The 4 \mj\ core-accretion planet, which is always relatively cool, is two magnitudes brighter in $M^{\prime}$ band than $L^{\prime}$.  However, the $M^{\prime}$-band flux will likely be depressed somewhat due to non-equilibrium CO/CH$_4$ chemistry \citep{Fegley96,Noll97,Saumon03,Saumon06}.  Recently \citet{Hubeny07} have investigated these affects for solar composition models across a wide range of \te, but at higher gravity than we consider here, and find a $\sim$40\% flux decrement in $M^{\prime}$ band due to non-equilibrium chemistry.  Interestingly, they find that this flux decrement increases with decreasing gravity (for a given $K_{zz}$, the eddy diffusion coefficient), meaning that non-equilibrium chemistry will likely remain important for young Jupiters.  Indeed, the effects of vertical mixing on CH$_4$/CO mixing ratios were first described in Jupiter itself \cp{Prinn77,Yung88,Fegley94}.  Although planets may well remain brighter in M$^{\prime}$ than L$^{\prime}$, the lower thermal background in L$^{\prime}$ for ground-based observatories will make both bands attractive for EGP searches.

While chemical reaction time constants can be measured (although there is uncertainly as to the actual reaction pathway for carbon chemistry), the mixing time scales are much more uncertain.  In the convective atmosphere the mixing time scale can be computed from mixing length theory.  However, in the radiative region, there is no \emph{a priori} theory to characterize this mixing, parametrized by $K_{zz}$.  Previous studies in brown dwarfs have varied this coefficient over 6-8 orders of magnitude.  Parametric studies of the effects of non-equilibrium chemistry in brown dwarfs, as as function of \te, gravity, and $K_{zz}$ can be found in \ct{Saumon03} and \ct{Hubeny07}.  In some instances $K_{zz}$ can be constrained by observations of brown dwarfs \cp[e.~g.][]{Saumon06}.  There is also a long history of modeling non-equilibrium chemistry in Jupiter, and recent estimates of $K_{zz}$ in Jupiter's radiative atmosphere range from $10^2$ to $10^4$ \cp{Bezard02,Moses05}.  There has not yet been a study of non-equilibrium chemistry at the low gravities relevant for young gas giants.  When abundant data for these planets becomes available, detailed studies will of course be necessary.  For now we will treat the effects of nonequilibrium chemistry as an uncertainty in the model spectra, which we can briefly gage.

\begin{figure}
\includegraphics[scale=0.75,angle=90]{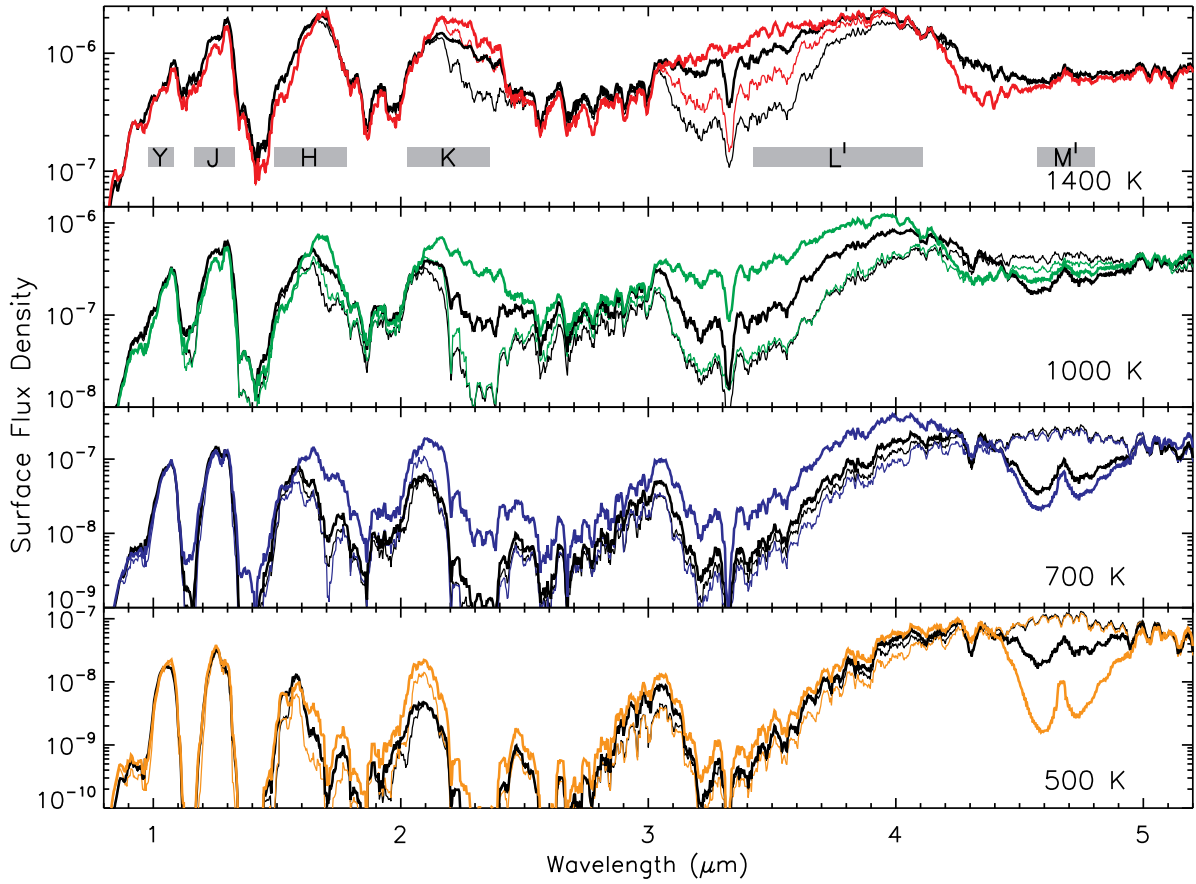}
\caption{Emergent spectra (erg s$^{-1}$ cm$^{-2}$ Hz$^{-1}$) for 5$\times$ solar ([M/H=0.7], colors) and 1$\times$ solar models ([M/H=0.0], black) at, from top to bottom, 1400, 1000, 700, and 500 K, for log $g$=3.67.  Thin lines are for equilibrium chemistry, as shown in \mbox{Figure~\ref{spec1}}.  Thick lines show models that utilize non-equilibrium chemistry with log $K_{zz}$=4.  Infrared filter bandpasses are shown in gray on the top panel.
\label{neq1}}
\end{figure}

Using the methods described in \ct{Saumon03,Saumon06,Saumon07} we have computed models with $K_{zz}=10^4\,$cm$^2$ s$^{-1}$, for 1$\times$ and 5$\times$ solar metallicity, which we show in \mbox{Figure~\ref{neq1}}.  These near-infrared spectra clearly show the effects of enhanced CO and depleted CH$_4$.  Absorption by CO in $M^{\prime}$ band remains strong down to 500 K, leading to a large flux decrement, especially in the 5$\times$ solar model.  In the $H$, $K$, and $L^{\prime}$ bands, the reduced CH$_4$ abundances leads to greater flux (sometimes dramatically) escaping through these bands.  The behavior as a function of gravity, metallicity, and $K_{zz}$ is complex due to a number of factors related to the regions of \emph{P-T} space that observations are sensitive to. For instance, lower gravity objects have lower pressure photospheres at a given \te, while higher atmospheric metallicity pushes the CO/CH$_4$ transition to higher pressure, while higher values of $K_{zz}$ lead to gas being mixed up from higher pressures and higher temperatures.

We caution that these nonequilibrium spectra should only be viewed as illustrative.  The non-equilibrium spectra are computed with the same \emph{P-T} structure as the equilibrium models, even though the gas composition and the opacity have changed. The resulting non-equilibrium spectra have a larger integrated flux than the corresponding equilibrium spectra.  In the future we will compute \emph{P-T} profiles that are consistent with the nonequilibrium chemical abundances, alleviating this problem \cp[e.~g.][]{Hubeny07}.

\section{Discussion \& Conclusions}
We have shown how metal enhanced atmospheres differ from their solar composition counterparts, in atmospheric structure, chemistry, spectra, and colors.  We have applied these results to an updated version of the \citet{Marley07} evolution models, which give cooling tracks for EGPs that are initially significantly colder than traditional models.  We urge caution in the application of the computed absolute magnitudes for these models provided in \mbox{Table~\ref{bigt5}}.  Recall that the \citet{Marley07} models incorporate the formation models of \citet{Hubickyj05}, which employ a treatment of accretion that is surely much simpler than what occurs in Nature.  The potential agreement or disagreement between observations and the model cooling tracks and magnitudes should not be taken as evidence for or against the viability of the core accretion formation scenario.  Indeed, the \citet{Hubickyj05} prescription is just one of several models of core accretion, which all currently include simplifications of the gas and solid accretion.  While we can claim with some confidence that young Jupiters are fainter than those predicted from an arbitrarily hot start, how much fainter depends sensitively on the details of accretion \cp{Marley07}.  Given the difficulty of predicting properties of EGPs at young ages, observations of these young objects will be of central importance.

\begin{figure}
\includegraphics[scale=0.75,angle=90]{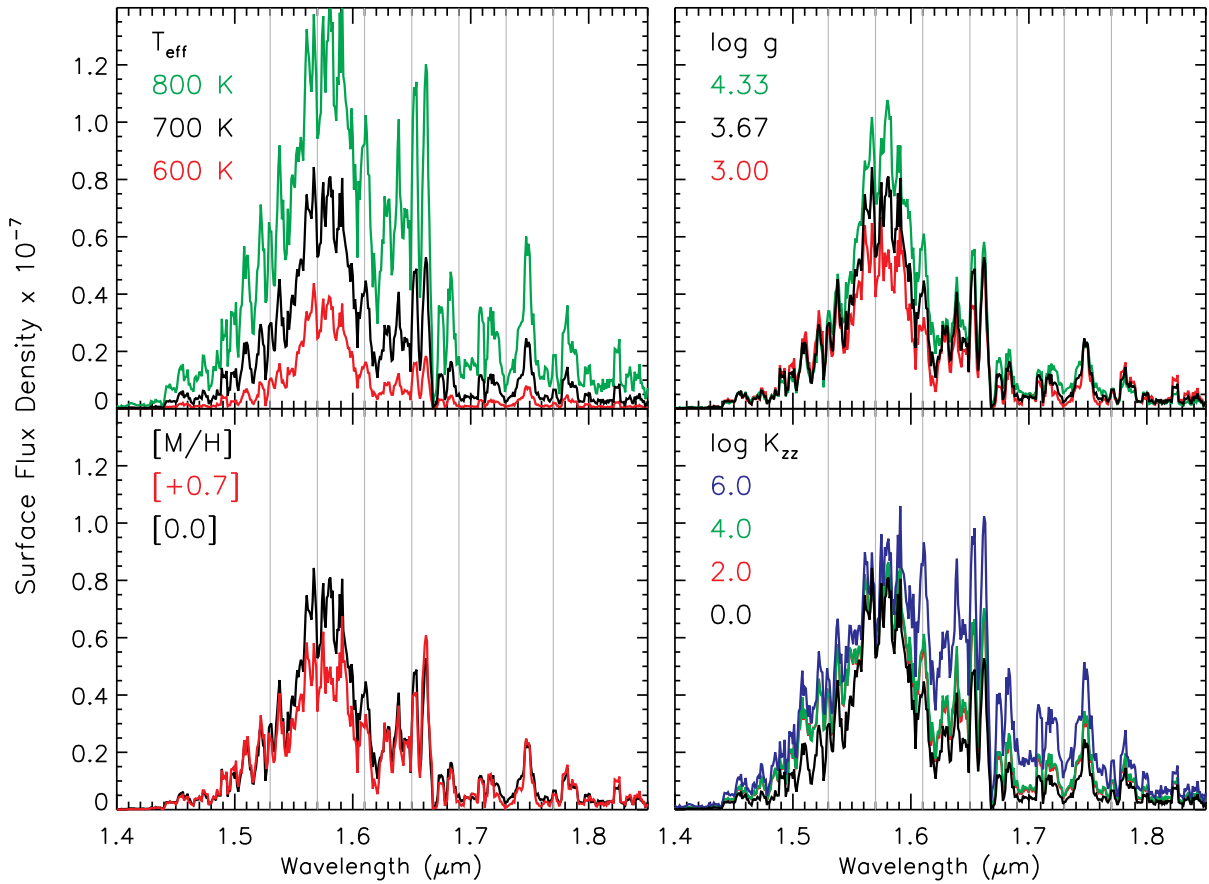}
\caption{A four-panel comparison of flux densities (erg s$^{-1}$ cm$^{-2}$ Hz$^{-1}$) around $H$-band.  All are referenced to a model with \te=700 K, log $g$=3.67, solar metallicity, and equilibrium chemistry.  The \te\ panel shows a $\pm$100 K change in \te.  The [M/H] panel compares the standard model to one that is 5$\times$ solar metallicity, [M/H=0.7].  The upper right panel (log g) shows the effects of gravity, while the lower right panel shows the effects of non-equilibrium chemistry due to vertical mixing.  Gray vertical lines guide the eye and illustrate possible locations of narrow band filters.
\label{pans}}
\end{figure}

The next generation of direct imaging platforms will be the Gemini Planet Imager (GPI), likely at Gemini South \cp{Macintosh06}, and Spectro-Polarimetric High-contrast Exoplanet REsearch (SPHERE), at the VLT \cp{Dohlen06}.  The GPI will target the $YJHK$ bands, while SPHERE will focus on $YJH$.  Both instruments will in particular emphasize $H$ band.  In \mbox{Figure~\ref{pans}} we show a four panel plot that illustrates changes to $H$-band fluxes due to \te, gravity, metallicity, and nonequilibrium chemistry.  All are referenced from a standard case model with \te=700 K.  Since GPI and SPHERE will likely employ custom narrow-band filters, we have overlain in gray possible narrow filters, kindly provided by J.~Graham.  It is clear that the more limited the observations are in wavelength, the more difficult planetary \emph{characterization} will be, as lower surface gravity and higher metallicity generally effect spectra in similar ways.  We must also caution that although our understanding of the gaseous opacity in these atmospheres is improving \cp{Sharp07,Freedman08}, calculations of the contribution due to CH$_4$ are quite uncertain.  The absorption cross-section of CH$_4$ is difficult to model under the relevant \emph{P-T} conditions found here, which manifests itself in mismatches of our models to brown dwarf spectra, especially around $H$-band \cp{Saumon06,Saumon07}.

Caveats now aside, we can readily summarize our findings for the low-gravity metal-enhanced young Jupiter atmospheres into the following four points:
\begin{enumerate}
\item Young Jupiter atmospheres will have lower pressure photospheres than old field brown dwarfs due to their lower surface gravity (which has long been understood) and higher atmospheric opacity (if the planets have high atmospheric metallicities, like Jupiter and Saturn).
\item Higher metallicity atmospheres, while generally having more opacity at all wavelengths, have relatively less opacity in $K$-band relative to other bands, due to weakened CIA H$_2$ opacity. This leads to a $K$-band flux enhancement of $\sim$0.5 to 1.0 magnitudes between \te s of 500 to 1400 K.
\item A spectral signature of high metallicity at \te $>$ 1000 K is a markedly deeper CO absorption band at 4.5 $\mu$m.
\item A photometric feature of high metallicity at \te $<$ 1400 K is redder $J-K$ and $H-K$ colors, which may be redder by $\sim$0.7-1.5.
\end{enumerate}
We note that points 1-3 in particular echo the findings of \ct{Chabrier07}, who had previously analyzed the spectrum of a representative metal-enhanced planet model within the parameter range we examine here.  The agreement is encouraging.

In closing, we note that the current best example of how well we may eventually be able to constrain the properties of a young EGP comes from the cool T7.5 dwarf Gliese 570D.  This brown dwarf is a wide companion to the well-studied K4 V star Gl 570A, and to a pair of M dwarfs, Gl 570BC.  As discussed by \citet{Saumon06}, the distance, metallicity, and age of the system effectively constrain the physical parameters of the T dwarf Gl 570D.  The spectrum is extremely well sampled from visible wavelengths, across the near infrared, to the mid infrared with \emph{Spitzer} IRS.  \ct{Saumon06} constrain \te =800-820 K, log $g$=5.09-5.23, log ($L$/$L_{\odot}$)=5.525-5.551, and mass=38-47 \mj.  \ct{Saumon07} additionally investigated two late field T dwarfs with similar spectral coverage, but without parent stars to constrain metallicities.  Uncertainties in \te\ increase $\sim$50-100 K, while mass estimates widen to $M \approx$30-60 \mj.  Since young Jupiters will have unknown metallicities and early on these objects will sample gravity and metallicity ranges that only marginally overlap the more well understood brown dwarfs, it will be challenging to constrain planetary parameters with limited photometric and spectral data.

\acknowledgements
We thank Bruce Macintosh and James Graham for numerous useful comments and suggestions, as well as for their enthusiasm for the project.  J.~J.~F.~acknowledges the support of a Spitzer Fellowship from NASA and NSF grant AST-0607489.  M.~S.~M.~acknowlodges the support of the NASA Planetary Atmospheres Program.  D.~S.~ acknowledges support from NASA through a Spitzer Space Telescope Grant through a contract issued by JPL/Caltech.  Work by K.~L.~is supported by NSF grant AST-0707377 and NASA grant NNG06GC26G.

%\bibliographystyle{apj}
%\bibliography{references}

\begin{thebibliography}{106}
\expandafter\ifx\csname natexlab\endcsname\relax\def\natexlab#1{#1}\fi

\bibitem[{{Ackerman} \& {Marley}(2001)}]{AM01}
{Ackerman}, A.~S. \& {Marley}, M.~S. 2001, \apj, 556, 872

\bibitem[{{Alibert} {et~al.}(2005{\natexlab{a}}){Alibert}, {Mordasini}, {Benz},
  \& {Winisdoerffer}}]{Alibert05}
{Alibert}, Y., {Mordasini}, C., {Benz}, W., \& {Winisdoerffer}, C.
  2005{\natexlab{a}}, \aap, 434, 343

\bibitem[{{Alibert} {et~al.}(2005{\natexlab{b}}){Alibert}, {Mousis},
  {Mordasini}, \& {Benz}}]{Alibert05b}
{Alibert}, Y., {Mousis}, O., {Mordasini}, C., \& {Benz}, W. 2005{\natexlab{b}},
  \apjl, 626, L57

\bibitem[{{Allard} {et~al.}(2001){Allard}, {Hauschildt}, {Alexander},
  {Tamanai}, \& {Schweitzer}}]{Allard01}
{Allard}, F., {Hauschildt}, P.~H., {Alexander}, D.~R., {Tamanai}, A., \&
  {Schweitzer}, A. 2001, \apj, 556, 357

\bibitem[{{Allard} {et~al.}(2003){Allard}, {Allard}, {Hauschildt}, {Kielkopf},
  \& {Machin}}]{Allard03}
{Allard}, N.~F., {Allard}, F., {Hauschildt}, P.~H., {Kielkopf}, J.~F., \&
  {Machin}, L. 2003, \aap, 411, L473

\bibitem[{{Apai} {et~al.}(2008){Apai}, {Janson}, {Moro-Mart{\'{\i}}n}, {Meyer},
  {Mamajek}, {Masciadri}, {Henning}, {Pascucci}, {Kim}, {Hillenbrand},
  {Kasper}, \& {Biller}}]{Apai08}
{Apai}, D., {Janson}, M., {Moro-Mart{\'{\i}}n}, A., {Meyer}, M.~R., {Mamajek},
  E.~E., {Masciadri}, E., {Henning}, T., {Pascucci}, I., {Kim}, J.~S.,
  {Hillenbrand}, L.~A., {Kasper}, M., \& {Biller}, B. 2008, \apj, 672, 1196

\bibitem[{{Atreya} {et~al.}(2003){Atreya}, {Mahaffy}, {Niemann}, {Wong}, \&
  {Owen}}]{Atreya03}
{Atreya}, S.~K., {Mahaffy}, P.~R., {Niemann}, H.~B., {Wong}, M.~H., \& {Owen},
  T.~C. 2003, \planss, 51, 105

\bibitem[{{Baraffe} {et~al.}(2002){Baraffe}, {Chabrier}, {Allard}, \&
  {Hauschildt}}]{Baraffe02}
{Baraffe}, I., {Chabrier}, G., {Allard}, F., \& {Hauschildt}, P.~H. 2002, A\&A,
  382, 563

\bibitem[{{Baraffe} {et~al.}(2008){Baraffe}, {Chabrier}, \&
  {Barman}}]{Baraffe08}
{Baraffe}, I., {Chabrier}, G., \& {Barman}, T. 2008, ArXiv e-prints/0802.1810

\bibitem[{{Baraffe} {et~al.}(2003){Baraffe}, {Chabrier}, {Barman}, {Allard}, \&
  {Hauschildt}}]{Baraffe03}
{Baraffe}, I., {Chabrier}, G., {Barman}, T.~S., {Allard}, F., \& {Hauschildt},
  P.~H. 2003, \aap, 402, 701

\bibitem[{{Beuzit} {et~al.}(2007){Beuzit}, {Mouillet}, {Oppenheimer}, \&
  {Monnier}}]{Beuzit07}
{Beuzit}, J.-L., {Mouillet}, D., {Oppenheimer}, B.~R., \& {Monnier}, J.~D.
  2007, in Protostars and Planets V, ed. B.~{Reipurth}, D.~{Jewitt}, \&
  K.~{Keil}, 717--732

\bibitem[{{B{\'e}zard} {et~al.}(2002){B{\'e}zard}, {Lellouch}, {Strobel},
  {Maillard}, \& {Drossart}}]{Bezard02}
{B{\'e}zard}, B., {Lellouch}, E., {Strobel}, D., {Maillard}, J.-P., \&
  {Drossart}, P. 2002, Icarus, 159, 95

\bibitem[{{Bodenheimer}(1976)}]{Bodenheimer76}
{Bodenheimer}, P. 1976, Icarus, 29, 165

\bibitem[{{Borysow}(2002)}]{Borysow02}
{Borysow}, A. 2002, \aap, 390, 779

\bibitem[{{Boss}(2001)}]{Boss01b}
{Boss}, A.~P. 2001, \apjl, 551, L167

\bibitem[{{Burrows} {et~al.}(2002){Burrows}, {Burgasser}, {Kirkpatrick},
  {Liebert}, {Milsom}, {Sudarsky}, \& {Hubeny}}]{Burrows02}
{Burrows}, A., {Burgasser}, A.~J., {Kirkpatrick}, J.~D., {Liebert}, J.,
  {Milsom}, J.~A., {Sudarsky}, D., \& {Hubeny}, I. 2002, \apj, 573, 394

\bibitem[{{Burrows} {et~al.}(2007){Burrows}, {Hubeny}, {Budaj}, \&
  {Hubbard}}]{Burrows07}
{Burrows}, A., {Hubeny}, I., {Budaj}, J., \& {Hubbard}, W.~B. 2007, \apj, 661,
  502

\bibitem[{{Burrows} {et~al.}(1997){Burrows}, {Marley}, {Hubbard}, {Lunine},
  {Guillot}, {Saumon}, {Freedman}, {Sudarsky}, \& {Sharp}}]{Burrows97}
{Burrows}, A., {Marley}, M., {Hubbard}, W.~B., {Lunine}, J.~I., {Guillot}, T.,
  {Saumon}, D., {Freedman}, R., {Sudarsky}, D., \& {Sharp}, C. 1997, \apj, 491,
  856

\bibitem[{{Burrows} {et~al.}(2000){Burrows}, {Marley}, \& {Sharp}}]{BMS}
{Burrows}, A., {Marley}, M.~S., \& {Sharp}, C.~M. 2000, \apj, 531, 438

\bibitem[{{Burrows} \& {Sharp}(1999)}]{Burrows99}
{Burrows}, A. \& {Sharp}, C.~M. 1999, \apj, 512, 843

\bibitem[{{Burrows} {et~al.}(2003){Burrows}, {Sudarsky}, \&
  {Lunine}}]{Burrows03b}
{Burrows}, A., {Sudarsky}, D., \& {Lunine}, J.~I. 2003, \apj, 596, 587

\bibitem[{{Chabrier} \& {Baraffe}(2000)}]{Chabrier00}
{Chabrier}, G. \& {Baraffe}, I. 2000, \araa, 38, 337

\bibitem[{{Chabrier} {et~al.}(2000){Chabrier}, {Baraffe}, {Allard}, \&
  {Hauschildt}}]{Chabrier00b}
{Chabrier}, G., {Baraffe}, I., {Allard}, F., \& {Hauschildt}, P. 2000, \apj,
  542, 464

\bibitem[{{Chabrier} {et~al.}(2007){Chabrier}, {Baraffe}, {Selsis}, {Barman},
  {Hennebelle}, \& {Alibert}}]{Chabrier07}
{Chabrier}, G., {Baraffe}, I., {Selsis}, F., {Barman}, T.~S., {Hennebelle}, P.,
  \& {Alibert}, Y. 2007, in Protostars and Planets V, ed. B.~{Reipurth},
  D.~{Jewitt}, \& K.~{Keil}, 623--638

\bibitem[{{Chauvin} {et~al.}(2005){Chauvin}, {Lagrange}, {Dumas}, {Zuckerman},
  {Mouillet}, {Song}, {Beuzit}, \& {Lowrance}}]{Chauvin05}
{Chauvin}, G., {Lagrange}, A.-M., {Dumas}, C., {Zuckerman}, B., {Mouillet}, D.,
  {Song}, I., {Beuzit}, J.-L., \& {Lowrance}, P. 2005, A\&A, 438, L25

\bibitem[{{Cooper} {et~al.}(2003){Cooper}, {Sudarsky}, {Milsom}, {Lunine}, \&
  {Burrows}}]{Cooper03}
{Cooper}, C.~S., {Sudarsky}, D., {Milsom}, J.~A., {Lunine}, J.~I., \&
  {Burrows}, A. 2003, \apj, 586, 1320

\bibitem[{{Cushing} {et~al.}(2007){Cushing}, {Marley}, {Saumon}, {Kelly},
  {Vacca}, {Rayner}, {Freedman}, {Lodders}, \& {Roellig}}]{Cushing08}
{Cushing}, M.~C., {Marley}, M.~S., {Saumon}, D., {Kelly}, B.~C., {Vacca},
  W.~D., {Rayner}, J.~T., {Freedman}, R.~S., {Lodders}, K., \& {Roellig}, T.~L.
  2007, ApJ in press, ArXiv e-prints/0711.0801

\bibitem[{{de Graauw} {et~al.}(1997){de Graauw}, {Feuchtgruber}, {Bezard},
  {Drossart}, {Encrenaz}, {Beintema}, {Griffin}, {Heras}, {Kessler}, {Leech},
  {Lellouch}, {Morris}, {Roelfsema}, {Roos-Serote}, {Salama}, {Vandenbussche},
  {Valentijn}, {Davis}, \& {Naylor}}]{deGraauw97}
{de Graauw}, T., {Feuchtgruber}, H., {Bezard}, B., {Drossart}, P., {Encrenaz},
  T., {Beintema}, D.~A., {Griffin}, M., {Heras}, A., {Kessler}, M., {Leech},
  K., {Lellouch}, E., {Morris}, P., {Roelfsema}, P.~R., {Roos-Serote}, M.,
  {Salama}, A., {Vandenbussche}, B., {Valentijn}, E.~A., {Davis}, G.~R., \&
  {Naylor}, D.~A. 1997, \aap, 321, L13

\bibitem[{{Dohlen} {et~al.}(2006){Dohlen}, {Beuzit}, {Feldt}, {Mouillet},
  {Puget}, {Antichi}, {Baruffolo}, {Baudoz}, {Berton}, {Boccaletti},
  {Carbillet}, {Charton}, {Claudi}, {Downing}, {Fabron}, {Feautrier},
  {Fedrigo}, {Fusco}, {Gach}, {Gratton}, {Hubin}, {Kasper}, {Langlois},
  {Longmore}, {Moutou}, {Petit}, {Pragt}, {Rabou}, {Rousset}, {Saisse},
  {Schmid}, {Stadler}, {Stamm}, {Turatto}, {Waters}, \& {Wildi}}]{Dohlen06}
{Dohlen}, K., {Beuzit}, J.-L., {Feldt}, M., {Mouillet}, D., {Puget}, P.,
  {Antichi}, J., {Baruffolo}, A., {Baudoz}, P., {Berton}, A., {Boccaletti}, A.,
  {Carbillet}, M., {Charton}, J., {Claudi}, R., {Downing}, M., {Fabron}, C.,
  {Feautrier}, P., {Fedrigo}, E., {Fusco}, T., {Gach}, J.-L., {Gratton}, R.,
  {Hubin}, N., {Kasper}, M., {Langlois}, M., {Longmore}, A., {Moutou}, C.,
  {Petit}, C., {Pragt}, J., {Rabou}, P., {Rousset}, G., {Saisse}, M., {Schmid},
  H.-M., {Stadler}, E., {Stamm}, D., {Turatto}, M., {Waters}, R., \& {Wildi},
  F. 2006, in Presented at the Society of Photo-Optical Instrumentation
  Engineers (SPIE) Conference, Vol. 6269, Ground-based and Airborne
  Instrumentation for Astronomy. Edited by McLean, Ian S.; Iye, Masanori.
  Proceedings of the SPIE, Volume 6269, pp. 62690Q (2006).

\bibitem[{{Encrenaz}(2005)}]{Encrenaz05}
{Encrenaz}, T. 2005, Space Science Reviews, 116, 99

\bibitem[{{Fegley} \& {Lodders}(1994)}]{Fegley94}
{Fegley}, B.~J. \& {Lodders}, K. 1994, Icarus, 110, 117

\bibitem[{{Fegley} \& {Lodders}(1996)}]{Fegley96}
---. 1996, \apjl, 472, L37

\bibitem[{{Flasar} {et~al.}(2005){Flasar}, {Achterberg}, {Conrath}, {Pearl},
  {Bjoraker}, {Jennings}, {Romani}, {Simon-Miller}, {Kunde}, {Nixon},
  {B{\'e}zard}, {Orton}, {Spilker}, {Spencer}, {Irwin}, {Teanby}, {Owen},
  {Brasunas}, {Segura}, {Carlson}, {Mamoutkine}, {Gierasch}, {Schinder},
  {Showalter}, {Ferrari}, {Barucci}, {Courtin}, {Coustenis}, {Fouchet},
  {Gautier}, {Lellouch}, {Marten}, {Prang{\'e}}, {Strobel}, {Calcutt}, {Read},
  {Taylor}, {Bowles}, {Samuelson}, {Abbas}, {Raulin}, {Ade}, {Edgington},
  {Pilorz}, {Wallis}, \& {Wishnow}}]{Flasar05}
{Flasar}, F.~M., {Achterberg}, R.~K., {Conrath}, B.~J., {Pearl}, J.~C.,
  {Bjoraker}, G.~L., {Jennings}, D.~E., {Romani}, P.~N., {Simon-Miller}, A.~A.,
  {Kunde}, V.~G., {Nixon}, C.~A., {B{\'e}zard}, B., {Orton}, G.~S., {Spilker},
  L.~J., {Spencer}, J.~R., {Irwin}, P.~G.~J., {Teanby}, N.~A., {Owen}, T.~C.,
  {Brasunas}, J., {Segura}, M.~E., {Carlson}, R.~C., {Mamoutkine}, A.,
  {Gierasch}, P.~J., {Schinder}, P.~J., {Showalter}, M.~R., {Ferrari}, C.,
  {Barucci}, A., {Courtin}, R., {Coustenis}, A., {Fouchet}, T., {Gautier}, D.,
  {Lellouch}, E., {Marten}, A., {Prang{\'e}}, R., {Strobel}, D.~F., {Calcutt},
  S.~B., {Read}, P.~L., {Taylor}, F.~W., {Bowles}, N., {Samuelson}, R.~E.,
  {Abbas}, M.~M., {Raulin}, F., {Ade}, P., {Edgington}, S., {Pilorz}, S.,
  {Wallis}, B., \& {Wishnow}, E.~H. 2005, Science, 307, 1247

\bibitem[{{Fortney} \& {Hubbard}(2003)}]{FH03}
{Fortney}, J.~J. \& {Hubbard}, W.~B. 2003, Icarus, 164, 228

\bibitem[{{Fortney} \& {Marley}(2007)}]{Fortney07b}
{Fortney}, J.~J. \& {Marley}, M.~S. 2007, \apjl, 666, L45

\bibitem[{{Fortney} {et~al.}(2007){Fortney}, {Marley}, \&
  {Barnes}}]{Fortney07a}
{Fortney}, J.~J., {Marley}, M.~S., \& {Barnes}, J.~W. 2007, \apj, 659, 1661

\bibitem[{{Fortney} {et~al.}(2005{\natexlab{a}}){Fortney}, {Marley},
  {Hubickyj}, {Bodenheimer}, \& {Lissauer}}]{Fortney05b}
{Fortney}, J.~J., {Marley}, M.~S., {Hubickyj}, O., {Bodenheimer}, P., \&
  {Lissauer}, J.~J. 2005{\natexlab{a}}, Astronomische Nachrichten, 326, 925

\bibitem[{{Fortney} {et~al.}(2005{\natexlab{b}}){Fortney}, {Marley}, {Lodders},
  {Saumon}, \& {Freedman}}]{Fortney05}
{Fortney}, J.~J., {Marley}, M.~S., {Lodders}, K., {Saumon}, D., \& {Freedman},
  R. 2005{\natexlab{b}}, \apjl, 627, L69

\bibitem[{{Fortney} {et~al.}(2006){Fortney}, {Saumon}, {Marley}, {Lodders}, \&
  {Freedman}}]{Fortney06}
{Fortney}, J.~J., {Saumon}, D., {Marley}, M.~S., {Lodders}, K., \& {Freedman},
  R.~S. 2006, \apj, 642, 495

\bibitem[{{Freedman} {et~al.}(2008){Freedman}, {Marley}, \&
  {Lodders}}]{Freedman08}
{Freedman}, R.~S., {Marley}, M.~S., \& {Lodders}, K. 2008, \apjs, 174, 504

\bibitem[{{Gautier} {et~al.}(2001{\natexlab{a}}){Gautier}, {Hersant}, {Mousis},
  \& {Lunine}}]{Gautier01a}
{Gautier}, D., {Hersant}, F., {Mousis}, O., \& {Lunine}, J.~I.
  2001{\natexlab{a}}, \apjl, 550, L227

\bibitem[{{Gautier} {et~al.}(2001{\natexlab{b}}){Gautier}, {Hersant}, {Mousis},
  \& {Lunine}}]{Gautier01b}
---. 2001{\natexlab{b}}, \apjl, 559, L183

\bibitem[{{Gautier} \& {Owen}(1989)}]{Gautier89}
{Gautier}, D. \& {Owen}, T. 1989, {in Origin and Evolution of Planetary and
  Satellite Atmospheres, ed. S. K. Ateya, J. B. Pollack, \& M. S. Matthews}
  (Tucson: Univ.~of Arizona Press), 487--512

\bibitem[{{Goody} {et~al.}(1989){Goody}, {West}, {Chen}, \& {Crisp}}]{Goody89}
{Goody}, R., {West}, R., {Chen}, L., \& {Crisp}, D. 1989, Journal of
  Quantitative Spectroscopy and Radiative Transfer, 42, 539

\bibitem[{{Graboske} {et~al.}(1975){Graboske}, {Olness}, {Pollack}, \&
  {Grossman}}]{Graboske75}
{Graboske}, H.~C., {Olness}, R.~J., {Pollack}, J.~B., \& {Grossman}, A.~S.
  1975, \apj, 199, 265

\bibitem[{{Guillot}(1999)}]{Guillot99}
{Guillot}, T. 1999, \planss, 47, 1183

\bibitem[{{Guillot} {et~al.}(1995){Guillot}, {Chabrier}, {Gautier}, \&
  {Morel}}]{Guillot95}
{Guillot}, T., {Chabrier}, G., {Gautier}, D., \& {Morel}, P. 1995, \apj, 450,
  463

\bibitem[{{Guillot} {et~al.}(1994){Guillot}, {Gautier}, {Chabrier}, \&
  {Mosser}}]{Guillot94a}
{Guillot}, T., {Gautier}, D., {Chabrier}, G., \& {Mosser}, B. 1994, Icarus,
  112, 337

\bibitem[{{Guillot} \& {Gladman}(2000)}]{Guillot00}
{Guillot}, T. \& {Gladman}, B. 2000, in ASP Conf. Ser. 219: Disks,
  Planetesimals, and Planets, ed. F. Garzon, C. Eiroa, D. de Winter, \& T. J.
  Mahoney (San Francisco: ASP), 475

\bibitem[{{Guillot} \& {Hueso}(2006)}]{Guillot06a}
{Guillot}, T. \& {Hueso}, R. 2006, \mnras, 367, L47

\bibitem[{{Guillot} {et~al.}(2006){Guillot}, {Santos}, {Pont}, {Iro}, {Melo},
  \& {Ribas}}]{Guillot06}
{Guillot}, T., {Santos}, N.~C., {Pont}, F., {Iro}, N., {Melo}, C., \& {Ribas},
  I. 2006, \aap, 453, L21

\bibitem[{{Guillot} {et~al.}(2004){Guillot}, {Stevenson}, {Hubbard}, \&
  {Saumon}}]{Guillot04}
{Guillot}, T., {Stevenson}, D.~J., {Hubbard}, W.~B., \& {Saumon}, D. 2004, {The
  interior of Jupiter} (Jupiter.~The Planet, Satellites and Magnetosphere),
  35--57

\bibitem[{{Hubbard}(1977)}]{Hubbard77}
{Hubbard}, W.~B. 1977, Icarus, 30, 305

\bibitem[{{Hubbard} \& {Marley}(1989)}]{Hubbard89}
{Hubbard}, W.~B. \& {Marley}, M.~S. 1989, Icarus, 78, 102

\bibitem[{{Hubeny} \& {Burrows}(2007)}]{Hubeny07}
{Hubeny}, I. \& {Burrows}, A. 2007, \apj, 669, 1248

\bibitem[{{Hubickyj} {et~al.}(2005){Hubickyj}, {Bodenheimer}, \&
  {Lissauer}}]{Hubickyj05}
{Hubickyj}, O., {Bodenheimer}, P., \& {Lissauer}, J.~J. 2005, Icarus, 179, 415

\bibitem[{{Ida} \& {Lin}(2004)}]{Ida04}
{Ida}, S. \& {Lin}, D.~N.~C. 2004, \apj, 604, 388

\bibitem[{{Ikoma} {et~al.}(2000){Ikoma}, {Nakazawa}, \& {Emori}}]{Ikoma00}
{Ikoma}, M., {Nakazawa}, K., \& {Emori}, H. 2000, \apj, 537, 1013

\bibitem[{{Kirkpatrick}(2007)}]{Kirkpatrick07}
{Kirkpatrick}, J.~D. 2007, Cool Stars 14, astro-ph/0704.1522

\bibitem[{{Knapp} {et~al.}(2004){Knapp}, {Leggett}, {Fan}, {Marley}, {Geballe},
  {Golimowski}, {Finkbeiner}, {Gunn}, {Hennawi}, {Ivezi{\'c}}, {Lupton},
  {Schlegel}, {Strauss}, {Tsvetanov}, {Chiu}, {Hoversten}, {Glazebrook},
  {Zheng}, {Hendrickson}, {Williams}, {Uomoto}, {Vrba}, {Henden}, {Luginbuhl},
  {Guetter}, {Munn}, {Canzian}, {Schneider}, \& {Brinkmann}}]{Knapp04}
{Knapp}, G.~R., {Leggett}, S.~K., {Fan}, X., {Marley}, M.~S., {Geballe}, T.~R.,
  {Golimowski}, D.~A., {Finkbeiner}, D., {Gunn}, J.~E., {Hennawi}, J.,
  {Ivezi{\'c}}, Z., {Lupton}, R.~H., {Schlegel}, D.~J., {Strauss}, M.~A.,
  {Tsvetanov}, Z.~I., {Chiu}, K., {Hoversten}, E.~A., {Glazebrook}, K.,
  {Zheng}, W., {Hendrickson}, M., {Williams}, C.~C., {Uomoto}, A., {Vrba},
  F.~J., {Henden}, A.~A., {Luginbuhl}, C.~B., {Guetter}, H.~H., {Munn}, J.~A.,
  {Canzian}, B., {Schneider}, D.~P., \& {Brinkmann}, J. 2004, \aj, 127, 3553

\bibitem[{{Kroupa} \& {Bouvier}(2003)}]{Kroupa03}
{Kroupa}, P. \& {Bouvier}, J. 2003, \mnras, 346, 369

\bibitem[{{Lafreni{\`e}re} {et~al.}(2007){Lafreni{\`e}re}, {Doyon}, {Marois},
  {Nadeau}, {Oppenheimer}, {Roche}, {Rigaut}, {Graham}, {Jayawardhana},
  {Johnstone}, {Kalas}, {Macintosh}, \& {Racine}}]{Lafreniere07}
{Lafreni{\`e}re}, D., {Doyon}, R., {Marois}, C., {Nadeau}, D., {Oppenheimer},
  B.~R., {Roche}, P.~F., {Rigaut}, F., {Graham}, J.~R., {Jayawardhana}, R.,
  {Johnstone}, D., {Kalas}, P.~G., {Macintosh}, B., \& {Racine}, R. 2007, \apj,
  670, 1367

\bibitem[{{Lissauer} \& {Stevenson}(2007)}]{Lissauer07}
{Lissauer}, J.~J. \& {Stevenson}, D.~J. 2007, in Protostars and Planets V, ed.
  B.~{Reipurth}, D.~{Jewitt}, \& K.~{Keil}, 591--606

\bibitem[{{Lodders}(1999)}]{Lodders99}
{Lodders}, K. 1999, \apj, 519, 793

\bibitem[{{Lodders}(2003)}]{Lodders03}
---. 2003, \apj, 591, 1220

\bibitem[{{Lodders}(2004)}]{Lodders04}
---. 2004, \apj, 611, 587

\bibitem[{{Lodders} \& {Fegley}(2002)}]{Lodders02}
{Lodders}, K. \& {Fegley}, B. 2002, Icarus, 155, 393

\bibitem[{{Lodders} \& {Fegley}(2006)}]{Lodders06}
---. 2006, {Astrophysics Update 2} (Springer Praxis Books, Berlin: Springer,
  2006)

\bibitem[{{Luhman} {et~al.}(2007{\natexlab{a}}){Luhman}, {Joergens}, {Lada},
  {Muzerolle}, {Pascucci}, \& {White}}]{Luhman07}
{Luhman}, K.~L., {Joergens}, V., {Lada}, C., {Muzerolle}, J., {Pascucci}, I.,
  \& {White}, R. 2007{\natexlab{a}}, in Protostars and Planets V, ed.
  B.~{Reipurth}, D.~{Jewitt}, \& K.~{Keil}, 443--457

\bibitem[{{Luhman} {et~al.}(2007{\natexlab{b}}){Luhman}, {Patten}, {Marengo},
  {Schuster}, {Hora}, {Ellis}, {Stauffer}, {Sonnett}, {Winston}, {Gutermuth},
  {Megeath}, {Backman}, {Henry}, {Werner}, \& {Fazio}}]{Luhman07b}
{Luhman}, K.~L., {Patten}, B.~M., {Marengo}, M., {Schuster}, M.~T., {Hora},
  J.~L., {Ellis}, R.~G., {Stauffer}, J.~R., {Sonnett}, S.~M., {Winston}, E.,
  {Gutermuth}, R.~A., {Megeath}, S.~T., {Backman}, D.~E., {Henry}, T.~J.,
  {Werner}, M.~W., \& {Fazio}, G.~G. 2007{\natexlab{b}}, \apj, 654, 570

\bibitem[{{Lunine} {et~al.}(1989){Lunine}, {Hubbard}, {Burrows}, {Wang}, \&
  {Garlow}}]{Lunine89}
{Lunine}, J.~I., {Hubbard}, W.~B., {Burrows}, A., {Wang}, Y.-P., \& {Garlow},
  K. 1989, \apj, 338, 314

\bibitem[{{Macintosh} {et~al.}(2006){Macintosh}, {Graham}, {Palmer}, {Doyon},
  {Gavel}, {Larkin}, {Oppenheimer}, {Saddlemyer}, {Wallace}, {Bauman}, {Evans},
  {Erikson}, {Morzinski}, {Phillion}, {Poyneer}, {Sivaramakrishnan}, {Soummer},
  {Thibault}, \& {Veran}}]{Macintosh06}
{Macintosh}, B., {Graham}, J., {Palmer}, D., {Doyon}, R., {Gavel}, D.,
  {Larkin}, J., {Oppenheimer}, B., {Saddlemyer}, L., {Wallace}, J.~K.,
  {Bauman}, B., {Evans}, J., {Erikson}, D., {Morzinski}, K., {Phillion}, D.,
  {Poyneer}, L., {Sivaramakrishnan}, A., {Soummer}, R., {Thibault}, S., \&
  {Veran}, J.-P. 2006, in Presented at the Society of Photo-Optical
  Instrumentation Engineers (SPIE) Conference, Vol. 6272, Advances in Adaptive
  Optics II. Edited by Ellerbroek, Brent L.; Bonaccini Calia, Domenico.
  Proceedings of the SPIE, Volume 6272, pp. 62720L (2006).

\bibitem[{{Marley} {et~al.}(2007{\natexlab{a}}){Marley}, {Fortney}, {Seager},
  \& {Barman}}]{Marley07b}
{Marley}, M.~S., {Fortney}, J., {Seager}, S., \& {Barman}, T.
  2007{\natexlab{a}}, in Protostars and Planets V, ed. B.~{Reipurth},
  D.~{Jewitt}, \& K.~{Keil}, 733--747

\bibitem[{{Marley} {et~al.}(2007{\natexlab{b}}){Marley}, {Fortney}, {Hubickyj},
  {Bodenheimer}, \& {Lissauer}}]{Marley07}
{Marley}, M.~S., {Fortney}, J.~J., {Hubickyj}, O., {Bodenheimer}, P., \&
  {Lissauer}, J.~J. 2007{\natexlab{b}}, \apj, 655, 541

\bibitem[{{Marley} \& {McKay}(1999)}]{MM99}
{Marley}, M.~S. \& {McKay}, C.~P. 1999, Icarus, 138, 268

\bibitem[{{Marley} {et~al.}(1996){Marley}, {Saumon}, {Guillot}, {Freedman},
  {Hubbard}, {Burrows}, \& {Lunine}}]{Marley96}
{Marley}, M.~S., {Saumon}, D., {Guillot}, T., {Freedman}, R.~S., {Hubbard},
  W.~B., {Burrows}, A., \& {Lunine}, J.~I. 1996, Science, 272, 1919

\bibitem[{{Marley} {et~al.}(2008){Marley}, {Saumon}, {Lodders}, {Freedman}, \&
  {Fortney}}]{Marley08}
{Marley}, M.~S., {Saumon}, D., {Lodders}, K., {Freedman}, R.~S., \& {Fortney},
  J.~J. 2008, in prep

\bibitem[{{Marley} {et~al.}(2002){Marley}, {Seager}, {Saumon}, {Lodders},
  {Ackerman}, {Freedman}, \& {Fan}}]{Marley02}
{Marley}, M.~S., {Seager}, S., {Saumon}, D., {Lodders}, K., {Ackerman}, A.~S.,
  {Freedman}, R.~S., \& {Fan}, X. 2002, \apj, 568, 335

\bibitem[{{McKay} {et~al.}(1989){McKay}, {Pollack}, \& {Courtin}}]{Mckay89}
{McKay}, C.~P., {Pollack}, J.~B., \& {Courtin}, R. 1989, Icarus, 80, 23

\bibitem[{{Metchev} \& {Hillenbrand}(2006)}]{Metchev06}
{Metchev}, S.~A. \& {Hillenbrand}, L.~A. 2006, \apj, 651, 1166

\bibitem[{{Moses} {et~al.}(2005){Moses}, {Fouchet}, {B{\'e}zard}, {Gladstone},
  {Lellouch}, \& {Feuchtgruber}}]{Moses05}
{Moses}, J.~I., {Fouchet}, T., {B{\'e}zard}, B., {Gladstone}, G.~R.,
  {Lellouch}, E., \& {Feuchtgruber}, H. 2005, Journal of Geophysical Research
  (Planets), 110, 8001

\bibitem[{{Nielsen} {et~al.}(2007){Nielsen}, {Close}, {Biller}, \&
  {Masciadri}}]{Nielsen07}
{Nielsen}, E.~L., {Close}, L.~M., {Biller}, B.~A., \& {Masciadri}, E. 2007, ApJ
  in press, astro-ph/0706.4331

\bibitem[{{Noll} {et~al.}(1997){Noll}, {Geballe}, \& {Marley}}]{Noll97}
{Noll}, K.~S., {Geballe}, T.~R., \& {Marley}, M.~S. 1997, \apjl, 489, L87

\bibitem[{{Owen} {et~al.}(1999){Owen}, {Mahaffy}, {Niemann}, {Atreya},
  {Donahue}, {Bar-Nun}, \& {de Pater}}]{Owen99}
{Owen}, T., {Mahaffy}, P., {Niemann}, H.~B., {Atreya}, S., {Donahue}, T.,
  {Bar-Nun}, A., \& {de Pater}, I. 1999, \nat, 402, 269

\bibitem[{{Podolak} \& {Cameron}(1974)}]{Podolak74}
{Podolak}, M. \& {Cameron}, A.~G.~W. 1974, Icarus, 22, 123

\bibitem[{{Pollack} {et~al.}(1996){Pollack}, {Hubickyj}, {Bodenheimer},
  {Lissauer}, {Podolak}, \& {Greenzweig}}]{Pollack96}
{Pollack}, J.~B., {Hubickyj}, O., {Bodenheimer}, P., {Lissauer}, J.~J.,
  {Podolak}, M., \& {Greenzweig}, Y. 1996, Icarus, 124, 62

\bibitem[{{Prinn} \& {Barshay}(1977)}]{Prinn77}
{Prinn}, R.~G. \& {Barshay}, S.~S. 1977, Science, 198, 1031

\bibitem[{{Ribas} \& {Miralda-Escud{\'e}}(2007)}]{Ribas07}
{Ribas}, I. \& {Miralda-Escud{\'e}}, J. 2007, \aap, 464, 779

\bibitem[{{Saumon} {et~al.}(1994){Saumon}, {Bergeron}, {Lunine}, {Hubbard}, \&
  {Burrows}}]{Saumon94}
{Saumon}, D., {Bergeron}, P., {Lunine}, J.~I., {Hubbard}, W.~B., \& {Burrows},
  A. 1994, \apj, 424, 333

\bibitem[{{Saumon} \& {Guillot}(2004)}]{Saumon04}
{Saumon}, D. \& {Guillot}, T. 2004, \apj, 609, 1170

\bibitem[{{Saumon} {et~al.}(1996){Saumon}, {Hubbard}, {Burrows}, {Guillot},
  {Lunine}, \& {Chabrier}}]{Saumon96}
{Saumon}, D., {Hubbard}, W.~B., {Burrows}, A., {Guillot}, T., {Lunine}, J.~I.,
  \& {Chabrier}, G. 1996, \apj, 460, 993

\bibitem[{{Saumon} \& {Marley}(2008)}]{Saumon08}
{Saumon}, D. \& {Marley}, M.~S. 2008, ApJ, submitted

\bibitem[{{Saumon} {et~al.}(2006){Saumon}, {Marley}, {Cushing}, {Leggett},
  {Roellig}, {Lodders}, \& {Freedman}}]{Saumon06}
{Saumon}, D., {Marley}, M.~S., {Cushing}, M.~C., {Leggett}, S.~K., {Roellig},
  T.~L., {Lodders}, K., \& {Freedman}, R.~S. 2006, \apj, 647, 552

\bibitem[{{Saumon} {et~al.}(2007){Saumon}, {Marley}, {Leggett}, {Geballe},
  {Stephens}, {Golimowski}, {Cushing}, {Fan}, {Rayner}, {Lodders}, \&
  {Freedman}}]{Saumon07}
{Saumon}, D., {Marley}, M.~S., {Leggett}, S.~K., {Geballe}, T.~R., {Stephens},
  D., {Golimowski}, D.~A., {Cushing}, M.~C., {Fan}, X., {Rayner}, J.~T.,
  {Lodders}, K., \& {Freedman}, R.~S. 2007, \apj, 656, 1136

\bibitem[{{Saumon} {et~al.}(2003){Saumon}, {Marley}, {Lodders}, \&
  {Freedman}}]{Saumon03}
{Saumon}, D., {Marley}, M.~S., {Lodders}, K., \& {Freedman}, R.~S. 2003, in
  Brown Dwarfs, Proceedings of IAU Symposium \#211, ed.~E.~Martin, (San
  Francisco: Astronomical Society of the Pacific), 345

\bibitem[{{Sharp} \& {Burrows}(2007)}]{Sharp07}
{Sharp}, C.~M. \& {Burrows}, A. 2007, \apjs, 168, 140

\bibitem[{{Showman} \& {Ingersoll}(1998)}]{Showman98}
{Showman}, A.~P. \& {Ingersoll}, A.~P. 1998, Icarus, 132, 205

\bibitem[{{Stahler} {et~al.}(1980){Stahler}, {Shu}, \& {Taam}}]{Stahler80}
{Stahler}, S.~W., {Shu}, F.~H., \& {Taam}, R.~E. 1980, \apj, 241, 637

\bibitem[{{Stevenson}(1982)}]{Stevenson82b}
{Stevenson}, D.~J. 1982, \planss, 30, 755

\bibitem[{{Stevenson}(1985)}]{Stevenson85}
---. 1985, Icarus, 62, 4

\bibitem[{{Stevenson} \& {Salpeter}(1977)}]{SS77b}
{Stevenson}, D.~J. \& {Salpeter}, E.~E. 1977, \apjs, 35, 239

\bibitem[{{Toon} {et~al.}(1989){Toon}, {McKay}, {Ackerman}, \&
  {Santhanam}}]{Toon89}
{Toon}, O.~B., {McKay}, C.~P., {Ackerman}, T.~P., \& {Santhanam}, K. 1989,
  Journal of Geophysical Research, 94, 16287

\bibitem[{{Trauger} \& {Traub}(2007)}]{Trauger07}
{Trauger}, J.~T. \& {Traub}, W.~A. 2007, \nat, 446, 771

\bibitem[{{Visscher} \& {Fegley}(2005)}]{Visscher05}
{Visscher}, C. \& {Fegley}, B.~J. 2005, \apj, 623, 1221

\bibitem[{{Whitworth} {et~al.}(2007){Whitworth}, {Bate}, {Nordlund},
  {Reipurth}, \& {Zinnecker}}]{Whitworth07}
{Whitworth}, A., {Bate}, M.~R., {Nordlund}, {\AA}., {Reipurth}, B., \&
  {Zinnecker}, H. 2007, in Protostars and Planets V, ed. B.~{Reipurth},
  D.~{Jewitt}, \& K.~{Keil}, 459--476

\bibitem[{{Yung} {et~al.}(1988){Yung}, {Drew}, {Pinto}, \& {Friedl}}]{Yung88}
{Yung}, Y.~L., {Drew}, W.~A., {Pinto}, J.~P., \& {Friedl}, R.~R. 1988, Icarus,
  73, 516

\end{thebibliography}

\clearpage

\begin{deluxetable}{ccccccccccc}
\tabletypesize{\scriptsize}
\center
\tablecolumns{11}
\tablewidth{0pc}
\tablecaption{1$\times$ Solar Absolute Magnitudes for ``Hot Start'' Evolution Models}
\tablehead{
\colhead{Mass (\mj)} & \colhead{t (yrs)} & \colhead{$T_{\rm eff}$ (K)} & \colhead{$R$ (\rj)} & \colhead{M$_Y$} & \colhead{M$_Z$} & \colhead{M$_J$} & \colhead{M$_H$} & \colhead{M$_K$} & \colhead{M$_{L'}$} & \colhead{M$_{M'}$}}
\startdata 

1.0 & 1.000e+06 & 900.3 & 1.57 & 16.20 & 16.97 & 15.05 & 14.44 & 14.01 & 12.52 & 11.22\\
. & 2.154e+06 & 747.5 & 1.53 & 17.28 & 18.13 & 16.09 & 15.67 & 15.14 & 13.49 & 11.58\\
. & 4.642e+06 & 644.2 & 1.43 & 17.82 & 18.80 & 16.75 & 16.87 & 16.37 & 14.31 & 12.04\\
. & 1.000e+07 & 554.6 & 1.35 & 18.22 & 19.35 & 17.41 & 18.27 & 17.87 & 15.19 & 12.63\\
2.0 & 1.000e+06 & 1266.7 & 1.72 & 14.36 & 15.00 & 13.18 & 12.41 & 11.72 & 10.43 & 10.45\\
. & 2.154e+06 & 1048.8 & 1.57 & 15.36 & 16.05 & 14.23 & 13.50 & 13.12 & 11.69 & 10.99\\
. & 4.642e+06 & 855.3 & 1.47 & 16.64 & 17.42 & 15.46 & 14.84 & 14.40 & 12.87 & 11.40\\
. & 1.000e+07 & 710.3 & 1.39 & 17.53 & 18.41 & 16.36 & 16.05 & 15.61 & 13.80 & 11.83\\
. & 2.154e+07 & 605.8 & 1.33 & 17.90 & 18.96 & 16.94 & 17.36 & 17.04 & 14.65 & 12.35\\
. & 4.642e+07 & 512.9 & 1.27 & 18.42 & 19.59 & 17.71 & 18.94 & 18.85 & 15.63 & 13.02\\
4.0 & 1.000e+06 & 1657.0 & 1.92 & 12.66 & 13.29 & 11.48 & 10.91 & 10.19 &  9.18 &  9.43\\
. & 2.154e+06 & 1432.0 & 1.71 & 13.73 & 14.35 & 12.53 & 11.84 & 11.09 &  9.92 & 10.15\\
. & 4.642e+06 & 1207.6 & 1.56 & 14.79 & 15.43 & 13.62 & 12.80 & 12.28 & 10.97 & 10.74\\
. & 1.000e+07 & 989.1 & 1.45 & 15.88 & 16.58 & 14.71 & 13.95 & 13.60 & 12.14 & 11.16\\
. & 2.154e+07 & 805.3 & 1.38 & 17.00 & 17.80 & 15.80 & 15.20 & 14.84 & 13.16 & 11.57\\
. & 4.642e+07 & 673.8 & 1.32 & 17.73 & 18.66 & 16.59 & 16.43 & 16.18 & 14.04 & 12.07\\
. & 1.000e+08 & 574.6 & 1.26 & 18.04 & 19.13 & 17.09 & 17.75 & 17.80 & 14.93 & 12.66\\
6.0 & 1.000e+06 & 1984.0 & 2.10 & 11.58 & 12.19 & 10.50 & 10.01 &  9.40 &  8.58 &  8.75\\
. & 2.154e+06 & 1720.9 & 1.82 & 12.63 & 13.25 & 11.46 & 10.89 & 10.22 &  9.25 &  9.45\\
. & 4.642e+06 & 1464.2 & 1.63 & 13.70 & 14.31 & 12.50 & 11.81 & 11.10 &  9.97 & 10.19\\
. & 1.000e+07 & 1219.6 & 1.50 & 14.83 & 15.45 & 13.65 & 12.82 & 12.32 & 11.04 & 10.79\\
. & 2.154e+07 & 988.4 & 1.40 & 16.02 & 16.72 & 14.81 & 14.03 & 13.68 & 12.21 & 11.22\\
. & 4.642e+07 & 803.5 & 1.33 & 17.06 & 17.86 & 15.80 & 15.25 & 14.91 & 13.19 & 11.63\\
. & 1.000e+08 & 674.9 & 1.28 & 17.81 & 18.73 & 16.58 & 16.45 & 16.25 & 14.04 & 12.13\\
. & 2.154e+08 & 574.9 & 1.22 & 18.12 & 19.19 & 17.07 & 17.75 & 17.89 & 14.94 & 12.73\\
8.0 & 1.000e+06 & 2184.9 & 2.28 & 10.91 & 11.48 &  9.92 &  9.44 &  8.89 &  8.17 &  8.30\\
. & 2.154e+06 & 1987.6 & 1.93 & 11.74 & 12.34 & 10.66 & 10.18 &  9.60 &  8.78 &  8.97\\
. & 4.642e+06 & 1656.0 & 1.69 & 12.91 & 13.52 & 11.72 & 11.14 & 10.47 &  9.48 &  9.70\\
. & 1.000e+07 & 1400.7 & 1.54 & 14.08 & 14.68 & 12.87 & 12.13 & 11.44 & 10.29 & 10.43\\
. & 2.154e+07 & 1163.4 & 1.42 & 15.15 & 15.78 & 13.97 & 13.14 & 12.71 & 11.38 & 10.95\\
. & 4.642e+07 & 940.0 & 1.34 & 16.39 & 17.10 & 15.13 & 14.39 & 14.03 & 12.51 & 11.37\\\
. & 1.000e+08 & 769.1 & 1.28 & 17.35 & 18.16 & 16.04 & 15.59 & 15.29 & 13.43 & 11.81\\
. & 2.154e+08 & 659.2 & 1.22 & 17.93 & 18.86 & 16.66 & 16.69 & 16.59 & 14.21 & 12.29\\
. & 4.642e+08 & 543.2 & 1.17 & 18.41 & 19.49 & 17.35 & 18.23 & 18.59 & 15.28 & 13.02\\
10.0 & 1.000e+06 & 2315.7 & 2.44 & 10.47 & 11.00 &  9.54 &  9.07 &  8.55 &  7.89 &  7.98\\
. & 2.154e+06 & 2168.4 & 2.04 & 11.18 & 11.75 & 10.18 &  9.71 &  9.17 &  8.45 &  8.61\\
. & 4.642e+06 & 1873.0 & 1.75 & 12.26 & 12.87 & 11.14 & 10.63 & 10.03 &  9.15 &  9.34\\
. & 1.000e+07 & 1553.8 & 1.57 & 13.34 & 13.94 & 12.15 & 11.52 & 10.85 &  9.80 & 10.06\\
. & 2.154e+07 & 1307.0 & 1.44 & 14.58 & 15.19 & 13.38 & 12.58 & 11.98 & 10.77 & 10.72\\
. & 4.642e+07 & 1072.7 & 1.35 & 15.68 & 16.33 & 14.47 & 13.66 & 13.30 & 11.89 & 11.17\\
. & 1.000e+08 & 871.5 & 1.28 & 16.83 & 17.58 & 15.54 & 14.88 & 14.55 & 12.89 & 11.57\\
. & 2.154e+08 & 735.9 & 1.22 & 17.60 & 18.45 & 16.29 & 15.95 & 15.76 & 13.68 & 12.02\\
. & 4.642e+08 & 614.5 & 1.16 & 18.07 & 19.08 & 16.88 & 17.24 & 17.43 & 14.60 & 12.61\\
. & 1.000e+09 & 498.2 & 1.12 & 18.88 & 19.97 & 17.84 & 18.82 & 19.51 & 15.67 & 13.36\\

\enddata
\tablecomments{Atmospheric metallicity is 1$\times$ solar.  MKO filter set used.  Time steps in years are equally spaced in log $t$.  Only models with \te$> 500$ are tabulated.}
\label{bigt1}
\end{deluxetable}

\newpage
\begin{deluxetable}{ccccccccccccc}
\tabletypesize{\scriptsize}
\center
\tablecolumns{11}
\tablewidth{0pc}
\tablecaption{5$\times$ Solar Absolute Magnitudes for Core-Accretion Start Evolution Models}
\tablehead{
\colhead{Mass (\mj)} & \colhead{t (yrs)} & \colhead{$T_{\rm eff}$ (K)} & \colhead{$R$ (\rj)} & \colhead{M$_I$} & \colhead{M$_Y$} & \colhead{M$_Z$} & \colhead{M$_J$} & \colhead{M$_H$} & \colhead{M$_K$} & \colhead{M$_{L'}$} & \colhead{M$_{M'}$} & \colhead{M$_{CH4}$}}
\startdata 
1.0 & 1.000e+06 & 672.5 & 1.46 & 21.70 & 16.75 & 18.03 & 16.16 & 17.07 & 16.15 & 14.63 & 12.19 & 16.23\\
. & 2.154e+06 & 641.6 & 1.43 & 22.03 & 17.00 & 18.31 & 16.44 & 17.50 & 16.52 & 14.89 & 12.33 & 16.59\\
. & 4.642e+06 & 592.9 & 1.38 & 22.58 & 17.43 & 18.76 & 16.91 & 18.22 & 17.12 & 15.32 & 12.58 & 17.21\\
. & 1.000e+07 & 528.6 & 1.32 & 23.51 & 18.13 & 19.48 & 17.72 & 19.43 & 18.07 & 15.94 & 12.98 & 18.30\\
. & 2.154e+07 & 455.8 & 1.26 & 24.76 & 19.21 & 20.56 & 18.95 & 21.05 & 19.40 & 16.81 & 13.55 & 19.78\\
2.0 & 1.000e+06 & 652.2 & 1.36 & 22.00 & 16.97 & 18.27 & 16.36 & 17.40 & 16.47 & 14.84 & 12.39 & 16.46\\
. & 2.154e+06 & 641.1 & 1.35 & 22.11 & 17.06 & 18.36 & 16.45 & 17.55 & 16.60 & 14.92 & 12.43 & 16.58\\
. & 4.642e+06 & 625.7 & 1.34 & 22.26 & 17.18 & 18.49 & 16.59 & 17.76 & 16.77 & 15.04 & 12.50 & 16.76\\
. & 1.000e+07 & 599.6 & 1.32 & 22.52 & 17.39 & 18.71 & 16.82 & 18.11 & 17.06 & 15.25 & 12.62 & 17.05\\
. & 2.154e+07 & 550.3 & 1.29 & 23.16 & 17.90 & 19.23 & 17.40 & 18.98 & 17.76 & 15.71 & 12.92 & 17.84\\
. & 4.642e+07 & 483.5 & 1.25 & 24.09 & 18.69 & 20.02 & 18.30 & 20.26 & 18.81 & 16.41 & 13.36 & 18.99\\
. & 1.000e+08 & 409.0 & 1.21 & 25.38 & 19.89 & 21.19 & 19.65 & 22.06 & 20.37 & 17.40 & 13.99 & 20.63\\
4.0 & 1.000e+06 & 585.3 & 1.26 & 22.71 & 17.57 & 18.87 & 16.97 & 18.32 & 17.29 & 15.36 & 12.78 & 17.19\\
. & 2.154e+06 & 584.2 & 1.26 & 22.73 & 17.58 & 18.88 & 16.98 & 18.34 & 17.30 & 15.37 & 12.78 & 17.20\\
. & 4.642e+06 & 580.6 & 1.26 & 22.77 & 17.62 & 18.92 & 17.02 & 18.40 & 17.35 & 15.41 & 12.81 & 17.26\\
. & 1.000e+07 & 573.1 & 1.26 & 22.86 & 17.69 & 19.00 & 17.11 & 18.52 & 17.46 & 15.48 & 12.85 & 17.37\\
. & 2.154e+07 & 558.8 & 1.25 & 23.03 & 17.84 & 19.14 & 17.27 & 18.76 & 17.66 & 15.61 & 12.94 & 17.58\\
. & 4.642e+07 & 536.7 & 1.24 & 23.29 & 18.06 & 19.37 & 17.52 & 19.13 & 17.97 & 15.82 & 13.07 & 17.91\\
. & 1.000e+08 & 492.8 & 1.22 & 23.83 & 18.54 & 19.84 & 18.05 & 19.90 & 18.63 & 16.26 & 13.35 & 18.60\\
. & 2.154e+08 & 428.1 & 1.19 & 24.84 & 19.52 & 20.80 & 19.14 & 21.43 & 20.02 & 17.12 & 13.89 & 19.98\\
6.0 & 1.000e+06 & 563.5 & 1.21 & 22.98 & 17.83 & 19.12 & 17.22 & 18.64 & 17.63 & 15.58 & 12.97 & 17.44\\
. & 2.154e+06 & 562.9 & 1.21 & 22.99 & 17.84 & 19.13 & 17.22 & 18.65 & 17.64 & 15.58 & 12.98 & 17.45\\
. & 4.642e+06 & 561.2 & 1.21 & 23.01 & 17.85 & 19.15 & 17.24 & 18.68 & 17.66 & 15.60 & 12.99 & 17.47\\
. & 1.000e+07 & 557.7 & 1.21 & 23.05 & 17.89 & 19.18 & 17.28 & 18.73 & 17.71 & 15.63 & 13.01 & 17.52\\
. & 2.154e+07 & 550.9 & 1.21 & 23.12 & 17.96 & 19.25 & 17.36 & 18.84 & 17.81 & 15.69 & 13.05 & 17.62\\
. & 4.642e+07 & 539.2 & 1.20 & 23.25 & 18.07 & 19.36 & 17.48 & 19.03 & 17.97 & 15.80 & 13.12 & 17.78\\
. & 1.000e+08 & 517.2 & 1.19 & 23.50 & 18.29 & 19.58 & 17.73 & 19.38 & 18.28 & 16.01 & 13.25 & 18.09\\
. & 2.154e+08 & 477.9 & 1.18 & 24.02 & 18.79 & 20.07 & 18.27 & 20.17 & 19.01 & 16.47 & 13.54 & 18.80\\
. & 4.642e+08 & 423.3 & 1.15 & 24.83 & 19.59 & 20.85 & 19.16 & 21.45 & 20.21 & 17.21 & 14.00 & 19.95\\
8.0 & 1.000e+06 & 556.3 & 1.17 & 23.09 & 17.95 & 19.23 & 17.32 & 18.74 & 17.78 & 15.67 & 13.08 & 17.50\\
. & 2.154e+06 & 556.0 & 1.17 & 23.09 & 17.95 & 19.24 & 17.32 & 18.74 & 17.78 & 15.67 & 13.08 & 17.50\\
. & 4.642e+06 & 555.2 & 1.17 & 23.10 & 17.96 & 19.24 & 17.33 & 18.75 & 17.80 & 15.68 & 13.09 & 17.51\\
. & 1.000e+07 & 553.2 & 1.17 & 23.12 & 17.98 & 19.26 & 17.35 & 18.79 & 17.82 & 15.70 & 13.10 & 17.54\\
. & 2.154e+07 & 549.4 & 1.17 & 23.16 & 18.02 & 19.30 & 17.39 & 18.85 & 17.88 & 15.73 & 13.12 & 17.59\\
. & 4.642e+07 & 541.2 & 1.17 & 23.25 & 18.10 & 19.38 & 17.48 & 18.97 & 17.99 & 15.81 & 13.17 & 17.71\\
. & 1.000e+08 & 524.4 & 1.16 & 23.43 & 18.26 & 19.55 & 17.66 & 19.24 & 18.23 & 15.97 & 13.27 & 17.94\\
. & 2.154e+08 & 496.5 & 1.15 & 23.74 & 18.55 & 19.83 & 17.98 & 19.70 & 18.65 & 16.24 & 13.45 & 18.35\\
. & 4.642e+08 & 462.1 & 1.13 & 24.23 & 19.05 & 20.31 & 18.52 & 20.50 & 19.42 & 16.71 & 13.74 & 19.07\\
10.0 & 1.000e+06 & 561.0 & 1.14 & 23.07 & 17.96 & 19.23 & 17.29 & 18.66 & 17.76 & 15.65 & 13.12 & 17.42\\
. & 2.154e+06 & 560.8 & 1.14 & 23.07 & 17.96 & 19.23 & 17.30 & 18.67 & 17.77 & 15.66 & 13.12 & 17.42\\
. & 4.642e+06 & 560.4 & 1.14 & 23.08 & 17.96 & 19.24 & 17.30 & 18.67 & 17.77 & 15.66 & 13.12 & 17.42\\
. & 1.000e+07 & 559.0 & 1.14 & 23.09 & 17.98 & 19.25 & 17.31 & 18.69 & 17.79 & 15.67 & 13.13 & 17.44\\
. & 2.154e+07 & 556.1 & 1.14 & 23.12 & 18.01 & 19.28 & 17.34 & 18.74 & 17.83 & 15.70 & 13.15 & 17.48\\
. & 4.642e+07 & 550.0 & 1.14 & 23.18 & 18.06 & 19.34 & 17.41 & 18.83 & 17.92 & 15.76 & 13.18 & 17.56\\
. & 1.000e+08 & 537.8 & 1.13 & 23.31 & 18.18 & 19.46 & 17.54 & 19.02 & 18.09 & 15.87 & 13.26 & 17.73\\
. & 2.154e+08 & 518.2 & 1.12 & 23.51 & 18.37 & 19.65 & 17.75 & 19.32 & 18.37 & 16.05 & 13.38 & 17.99\\
. & 4.642e+08 & 489.7 & 1.11 & 23.84 & 18.69 & 19.96 & 18.10 & 19.83 & 18.86 & 16.36 & 13.57 & 18.45\\
. & 1.000e+09 & 442.1 & 1.09 & 24.49 & 19.36 & 20.61 & 18.83 & 20.93 & 19.94 & 17.01 & 13.98 & 19.43\\

\enddata
\tablecomments{Atmospheric metallicity is 5$\times$ solar.  MKO filter set used.  Filter ``CH4'' is from 1.57-1.61 $\mu$m, near a peak in planetary emission, as shown in \mbox{Figure~\ref{pans}}.  The state of the planet at the end of the core-accretion method of formation, from \ct{Hubickyj05}, is highly uncertain, so these predictions should used with care.  See \ct{Marley07} for a discussion.}
\label{bigt5}
\end{deluxetable}

\clearpage

\end{document}